\documentclass[twocolumn]{aastex631}

\usepackage{amsmath}
\usepackage{amssymb}

\usepackage{xcolor}
\definecolor{darkpurple}{HTML}{561977}

\shorttitle{RML for ALMA Protoplanetary Disks}
\shortauthors{Zawadzki et al.}

\begin{document}

\title{Regularized Maximum Likelihood Image Synthesis and Validation for ALMA Continuum Observations of Protoplanetary Disks}

\author[0000-0001-9319-1296]{Brianna Zawadzki}
\affiliation{Department of Astronomy and Astrophysics, 525 Davey Laboratory, The Pennsylvania State University, University Park, PA 16802, USA}
\affiliation{Center for Exoplanets and Habitable Worlds, 525 Davey Laboratory, The Pennsylvania State University, University Park, PA 16802, USA}
\affiliation{Institute for Computational \& Data Sciences, The Pennsylvania State University, University Park, PA 16802, USA}

\author[0000-0002-1483-8811]{Ian Czekala}
\affiliation{Department of Astronomy and Astrophysics, 525 Davey Laboratory, The Pennsylvania State University, University Park, PA 16802, USA}
\affiliation{Center for Exoplanets and Habitable Worlds, 525 Davey Laboratory, The Pennsylvania State University, University Park, PA 16802, USA}
\affiliation{Center for Astrostatistics, 525 Davey Laboratory, The Pennsylvania State University, University Park, PA 16802, USA}
\affiliation{Institute for Computational \& Data Sciences, The Pennsylvania State University, University Park, PA 16802, USA}

\author[0000-0002-8932-1219]{Ryan A. Loomis}
\affiliation{National Radio Astronomy Observatory, 520 Edgemont Rd., Charlottesville, VA 22903, USA}

\author[0000-0002-8974-8095]{Tyler Quinn}
\affiliation{Department of Astronomy and Astrophysics, 525 Davey Laboratory, The Pennsylvania State University, University Park, PA 16802, USA}

\author{Hannah Grzybowski}
\affiliation{Department of Astronomy and Astrophysics, 525 Davey Laboratory, The Pennsylvania State University, University Park, PA 16802, USA}

\author{Robert C. Frazier}
\affiliation{Department of Astronomy and Astrophysics, 525 Davey Laboratory, The Pennsylvania State University, University Park, PA 16802, USA}

\author[0000-0002-7032-2350]{Jeff Jennings}
\affiliation{Department of Astronomy and Astrophysics, 525 Davey Laboratory, The Pennsylvania State University, University Park, PA 16802, USA}

\author[0000-0002-7217-446X]{ Kadri M. Nizam}
\affiliation{Department of Astronomy and Astrophysics, 525 Davey Laboratory, The Pennsylvania State University, University Park, PA 16802, USA}
\affiliation{Center for Exoplanets and Habitable Worlds, 525 Davey Laboratory, The Pennsylvania State University, University Park, PA 16802, USA}
\affiliation{Center for Astrostatistics, 525 Davey Laboratory, The Pennsylvania State University, University Park, PA 16802, USA}
\affiliation{Institute for Computational \& Data Sciences, The Pennsylvania State University, University Park, PA 16802, USA}

\author{Yina Jian}
\affiliation{National Radio Astronomy Observatory, 520 Edgemont Rd., Charlottesville, VA 22903, USA}

\begin{abstract}

Regularized Maximum Likelihood (RML) techniques are a class of image synthesis methods that achieve better angular resolution and image fidelity than traditional methods like CLEAN for sub-mm interferometric observations. To identify best practices for RML imaging, we used the GPU-accelerated open source Python package \texttt{MPoL}, a machine learning-based RML approach, to explore the influence of common RML regularizers (maximum entropy, sparsity, total variation, and total squared variation) on images reconstructed from real and synthetic ALMA continuum observations of protoplanetary disks. We tested two different cross-validation (CV) procedures to characterize their performance and determine optimal prior strengths, and found that CV over a coarse grid of regularization strengths easily identifies a range of models with comparably strong predictive power. To evaluate the performance of RML techniques against a ground truth image, we used \texttt{MPoL} on a synthetic protoplanetary disk dataset and found that RML methods successfully resolve structures at fine spatial scales present in the original simulation. We used ALMA DSHARP observations of the protoplanetary disk around HD~143006 to compare the performance of \texttt{MPoL} and CLEAN, finding that RML imaging improved the spatial resolution of the image by up to a factor of $3$ without sacrificing sensitivity. We provide general recommendations for building an RML workflow for image synthesis of ALMA protoplanetary disk observations, including effective use of CV. Using these techniques to improve the imaging resolution of protoplanetary disk observations will enable new science, including the detection of protoplanets embedded in disks.
\end{abstract}


\section{Introduction} \label{sec:intro}

Sub-mm interferometric observations of protoplanetary disks provide critical insight into disk properties like temperatures and densities which can be used to better understand the planet formation process. Observations of disks have supported theoretical models of grain growth, planetesimal and planet formation, and the emergence of disk substructures \citep[e.g.][]{Isella_2010, Isella_2016, Perez_2012, ALMA_2015, Perez_2015, Andrews_2016, Cieza_2016, Cieza_2017, Zhang_2016, Pinte_2018, Tripathi_2018}. In recent years, interferometric observations that achieve high angular resolution \citep[some down to scales of 35 mas (5 AU), such as DSHARP; see][]{Andrews_2018, Huang_2018} have contributed to a growing catalog of highly resolved protoplanetary disks. Making further progress requires accurately imaging disks at still finer spatial scales. As these scales are resolved, our ability to characterize dust and gas disk substructures will improve, including the ability to detect signatures of forming planets embedded within their disks \citep{Benisty_2021, Casassus_2021, Bae_2022}. 

The Atacama Large Millimeter/submillimeter Array (ALMA) is capable of observing sources at high angular and spectral resolution, down to $20$~mas angular resolution at an observing wavelength of $1.3$~mm ($230$~GHz) \citep{ALMA_2022}. Interferometers such as ALMA are composed of a number of individual antennas, with every pair defining a baseline. Because of practical limitations in the number and placement of antennas as well as observation duration, only a finite subset of baseline lengths are sampled during an observation. As a result, interferometers incompletely and noisily sample the visibility function of an astronomical source, given by
\begin{equation}\label{eqn:vis}
    {\cal V}(u,v) = \iint I(l,m) \exp \left \{- 2 \pi i (ul + vm) \right \} \mathrm{d}l\,\mathrm{d}m.
\end{equation}
Here, ${\cal V}(u,v)$ is the visibility function parameterized by spatial frequencies $u$ and $v$, and $I(l,m)$ is the sky brightness distribution, where $l=\sin (\Delta \alpha \cos \delta)$ and $m=\sin (\Delta \delta)$ for right ascension $\alpha$ and declination $\delta$. The visibility function ${\cal V}$ and the sky brightness distribution $I$ are related by the Fourier transform, with the primary data product from the interferometer being a set of $k$ visibility measurements $\mathbf{D}$ at Fourier domain coordinates $(u_{k},v_{k})$. 

Interferometric images are synthesized from the observed visibility data --- the final image product depends on how the algorithm treats noisy visibility measurements and what assumptions are made about the unsampled spatial frequencies. If all unsampled spatial frequencies are set to zero power, the inverse Fourier transform of the visibilities under the chosen weighting scheme delivers the \textit{dirty image}. The dirty image can be thought of as a convolution of the true sky brightness distribution and the instrument point spread function (PSF), or dirty beam \citep{Hoegbom_1974}. Because beam sidelobes add image artifacts that are not representative of the true source sky brightness, dirty images require processing to better reconstruct the sky brightness. A more detailed overview of the general imaging process is provided in \citet[Ch. 10,11;][]{Thompson_2017}.

\subsection{Image Synthesis with CLEAN}
CLEAN is currently one of the most popular and well-supported imaging methods in radio interferometric software \citep{McMullin_2007, CASA_2022}. CLEAN begins with the dirty image and iteratively \lq{}deconvolves\rq{} beam sidelobes while building up a model representation of the sky brightness \citep{Hoegbom_1974}. The model is built from CLEAN components, usually Dirac $\delta$-functions or two-dimensional Gaussian components, which are placed at the location of the brightest pixel at each iteration. Deconvolution occurs when the CLEAN component is convolved with the dirty beam and subtracted from the dirty image. This process repeats until either a certain number of iterations is reached or the dirty image reaches some noise threshold. There are two products at the end of the CLEANing process: the residual image (originally the dirty image, but now contains only residuals after deconvolution) and the CLEAN model (composed of CLEAN components). The CLEAN model is then convolved with the CLEAN beam (usually a Gaussian fit to the main lobe of the dirty beam) and added to the residual image to form the final CLEANed image \citep[Ch. 11.1;][]{Thompson_2017}.

Although CLEAN has long been a reliable way to process images, it has limitations. Standard CLEAN components are simplistic (e.g. a Gaussian), which may not be suitable for capturing certain morphologies, such as sharp edges or rings in a disk. Extensions to CLEAN, such as adaptive- or multi-scale approaches that use different component sizes, can yield an image that is a more realistic representation of an extended source, though it may still be difficult to accurately reconstruct all features \citep{Bhatnagar_2004, Cornwell_2008}. Regardless of the chosen variant of CLEAN, it is common practice to convolve the CLEAN model with a final restoring beam. This convolution makes the CLEANed image more visually pleasing, but acts as a low pass filter, spatially broadening all information (most strongly that at high resolution) in the CLEAN model. Convolution with the CLEAN beam thus imposes a resolution limit on the final image. The resolution and sensitivity of a CLEAN image also depend on how the visibilities are weighted; uniform weighting results in images with high resolution and low sensitivity, while natural weighting favors sensitivity at the cost of resolution. Robust weighting allows for an adjustable resolution-sensitivity trade-off by selecting a robust parameter $-2 \le R \le 2$, where $R=-2$ is similar to uniform weighting and $R=2$ is similar to natural weighting \citep{Briggs_1995}.

Other drawbacks of CLEAN include the computational speed; CLEANing even a single pointing image cube can take several hours, while other image synthesis procedures developed with more modern computational infrastructure in mind can often synthesize an image at least an order of magnitude faster \citep[e.g.][]{Carcamo_2018}. Lastly, CLEAN is a nonlinear image restoration procedure rather than a true optimization algorithm; at least in the Common Astronomy Software Applications (CASA) \texttt{tclean} implementation, there are many user-specified algorithm parameters that could affect the outcome of the CLEANing process, e.g. stopping criteria and masks that limit where CLEAN components may be placed \citep{McMullin_2007}. Many of these parameters do not have a clear best choice that corresponds with the image qualities needed or desired for a given science case, nor can they be determined a priori. Rather, these parameters need to be determined by experimentation, which can be laborious when parameters interact strongly with each other.

\subsection{Alternative Image Synthesis Methods}

Many of the drawbacks encountered with CLEAN can be partially or completely avoided by using an alternative class of imaging techniques which incorporate additional information into the image synthesis routine through the use of regularizers. These methods have been successful across a diverse array of methodologies and implementations, including maximum entropy methods \citep[MEM, e.g.][]{Posonby_1973, Ables_1974, Cornwell_1985, Narayan_1986, Casassus_2013}, compressed sensing and sparse reconstruction methods \citep[e.g.][]{Wiaux_2009, Li_2011, Dabbech_2015, Onose_2016}, visibility model fitting \citep[e.g.][]{Tazzari_2018, Jennings_2020}, or machine learning-based methods \citep[e.g.][]{SanchezBermudez_2022, Terris_2023, Dabbech_2022, Veneri_2023}.

In general, regularized maximum likelihood (RML) imaging refers to image synthesis methods that require maximizing the likelihood of a set of visibility data, given a set of predicted model visibility values and regularizers. RML imaging techniques have applications in optical \citep[e.g.][]{Buscher_1994, Thiebaut_2008, Claes_2020}, infrared \citep[e.g.][]{Baron_2010}, and radio interferometry \citep[e.g.][]{Narayan_1986, EHTCollab_2019}. These techniques can include well-known regularizers like MEM, but numerous other ways to make assumptions about the source via regularization also exist.

A notable example in sub-mm radio interferometry are the Event Horizon Telescope images synthesized from observations of M87. The team successfully used two independently-developed RML pipelines to obtain high-resolution images of M87 and showed that RML methods can produce higher resolution images than CLEAN at similar image fidelity requirements \citep{Chael_2018, EHTCollab_2019}. For ALMA continuum observations of protoplanetary disks, \citet{Carcamo_2018} and \citet{Perez_2019} successfully used RML imaging with entropy-based regularizers on observations of HL~Tau and HD~169142 respectively, finding that RML methods can not only achieve better resolution than the corresponding CASA \texttt{tclean} image, but also suppress background noise more effectively. In addition, \citet{Yamaguchi_2020} applied RML imaging techniques with sparsity and total squared variation regularizers to ALMA observations of the protoplanetary disk around HD~142527, yielding images with improved fidelity and higher angular resolution compared to their CLEAN counterparts.

Despite these notable and impressive applications of applying RML imaging techniques to ALMA observations of protoplanetary disks thus far, there has not yet been a systematic exploration to test the imaging outcomes of various regularizers on ALMA protoplanetary disk observations, nor has there been an analysis of image validation procedures and regularizer tuning for these datasets. In this paper we explore the effects of four different regularizers (entropy, sparsity, total variation, and total squared variation) on both real and simulated ALMA continuum observations of protoplanetary disks and examine the images resulting from different image validation methods. We describe the data used in this study and how it was prepared for RML imaging in section \ref{sec:data}. In section \ref{sec:methods} we discuss the theory of RML imaging, including the mathematical forms of various regularizers; give a technical overview of the RML imaging Python package \texttt{MPoL}; and describe image validation procedures. We discuss the behavior and attributes of different regularizing terms in section \ref{sec:results}, followed by a more thorough examination of image validation procedures and a characterization of RML image resolution in section \ref{sec:discussion}. We present our conclusions in section \ref{sec:conclusion}. Appendix \ref{app:recs} contains recommendations for developing a successful RML workflow for ALMA measurement sets of protoplanetary disk observations.

\section{Data} 
\label{sec:data}

Throughout this study, we used three reference ALMA visibility datasets. The first is a small, mock dataset created from the ALMA logo (containing only $\sim1\%$ of the number of visibilities of a real, high-resolution ALMA dataset). Its small size means that it can be easily stored and processed on servers with limited computational means. A full accounting of the data processing steps are available on the \texttt{mpoldatasets}  repository;\footnote{\url{https://github.com/MPoL-dev/mpoldatasets}} we briefly summarize them here. The logo was converted to a grayscale image, Fourier transformed, apodized with a Blackman Harris window function (to remove spatial frequencies substantially higher than will be sampled by the target array), and saved as a FITS file. We then used the CASA task \texttt{simobserve} \citep[CASA version 6.1;][]{McMullin_2007} with the C43-7 reference ALMA configuration from simobserve (\texttt{alma.cycle7.7.cfg}), to ``observe'' the source as it transits zenith for 1 hour under median atmospheric conditions.

The second dataset in this study is a real Band 6 ALMA dataset containing the observations of the protoplanetary disk hosted by HD~143006, obtained by the DSHARP survey \citep{Andrews_2018} at a resolution of 45 mas. We chose this protoplanetary disk because it is well-studied and has potential for structures at small spatial scales, including azimuthal asymmetries. The visibilities were originally calibrated by the DSHARP team following the standardized CASA procedures described in \citet{Andrews_2018}. The full listing of archival observations can be found in \citet[][Table 3]{Andrews_2018}, and the calibrated visibilities can be downloaded from the DSHARP archive\footnote{\url{https://almascience.eso.org/almadata/lp/DSHARP/MSfiles/HD143006_continuum.ms.tgz}}. We performed one additional step of calibration beyond that of the DSHARP team. We found that the definition of the visibility weights was not consistent across all of the archival datasets, most likely because the treatment of statistical weights used to calibrate the visibilities frequently changed in 4.x versions of CASA. We found empirical weight scalings for each dataset by creating a \texttt{tclean} model, subtracting it from the visibilities, and examining the scatter in the visibility residuals compared to the Gaussian envelope expected from the thermal weights ($w=\sigma^{-2}$). Each spectral window was corrected individually by multiplying $\sigma$ by a scale factor for that spectral window, with a minimum scale factor of 1.46, a maximum of 1.91, and an average of 1.75 over all spectral windows. A walkthrough of this rescaling process is described as part of the \texttt{MPoL} documentation\footnote{\url{https://mpol-dev.github.io/visread/tutorials/rescale_AS209_weights.html}} and is documented in the \texttt{mpoldatasets} repository\footnote{\url{https://github.com/MPoL-dev/mpoldatasets/tree/main/products/HD143006-DSHARP-continuum}}.

The third dataset is a synthetic ALMA dataset we generated from a protoplanetary disk simulation described in \citet{Pinte_2016}. We converted the model image to grayscale and apodized the edges using a Hann window function. We scaled the simulated image to $512 \times 512$ pixels, with each pixel measuring 0.01 arcsec across so that the total angular extent of the emission was comparable to disks in the DSHARP survey. Using the Python Imaging Library, we scaled the total flux of the image to 59 mJy, matching the total flux of HD~143006 \citep{Andrews_2018}. We applied the \texttt{MPoL} routine for synthetic data generation, which uses a non-uniform fast Fourier transform (NuFFT) to calculate model visibilities at specified $(u,v)$. We used the same $(u,v)$ sampling of the DSHARP observations of HD~143006. We then added random Gaussian noise to the complex visibilities, with the noise amplitude distribution set by the inverse square-root of the weights. The purpose of this dataset is to compare RML model results with a realistic reference image which we can use as a ``ground truth.''

Before performing any imaging (either creating a dirty image or an RML image), we take the ungridded visibility data and average it to grid cells in the visibility domain. We specify the grid cells by first defining the spatial extent and desired number of pixels of the image. Then, we define a corresponding Fourier grid with the same number of grid cells as the image has pixels. The ungridded visibilities can be averaged using a simple weighted average, which is equivalent to uniform weighting. RML images in \texttt{MPoL} always begin with uniformly weighted visibilities, as only uniform weighting retains the statistical properties of the data needed for forward modeling.


\section{Forward Modeling with RML} \label{sec:methods}

\begin{figure*}
\centering
\includegraphics[width=\linewidth]{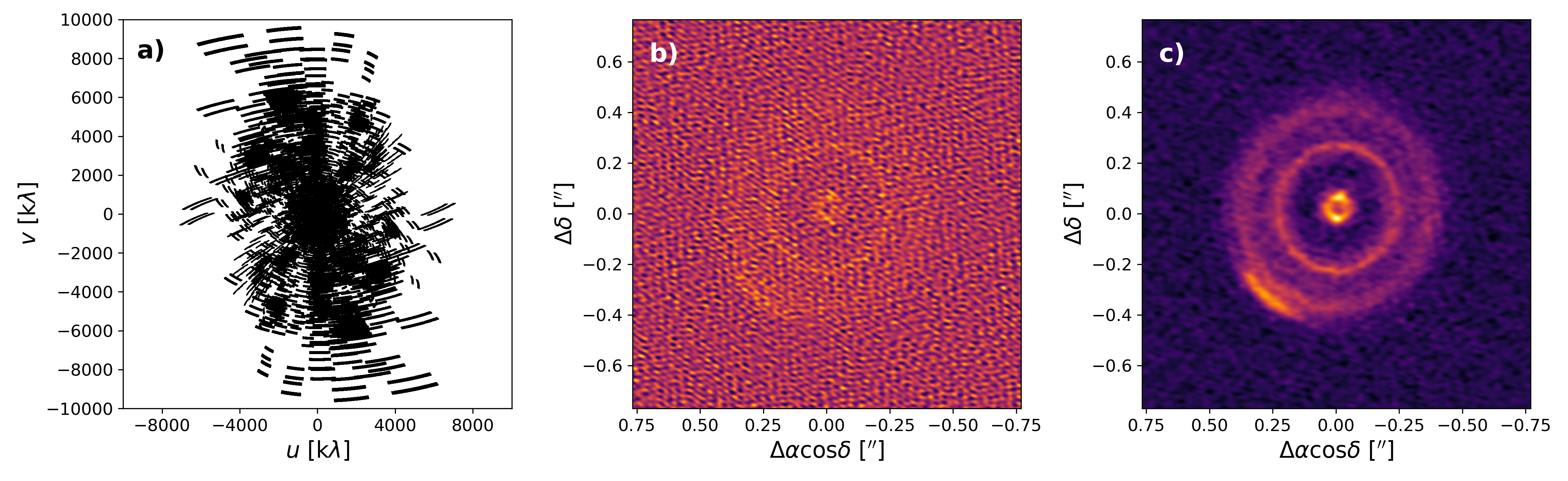}
\caption{An example of how incomplete $(u,v)$ sampling can impact the appearance of the dirty image. a) The $(u,v)$ sampling of the HD~143006 dataset. Even though the dataset contains many long-baseline visibilities, they are fewer in number than short-baseline visibilities. Gaps and low-sensitivity regions of $(u,v)$ space at a variety of baselines create artifacts in the dirty images because the power at these unsampled spatial frequencies is assumed to be zero. b) The dirty image with uniform weighting, which favors resolution over sensitivity. c) The dirty image with Briggs weighting with a robust parameter of 0.0, which creates more balance between resolution and sensitivity but still requires a tradeoff. Both dirty images were made with the \texttt{MPoL} \texttt{DirtyImager} module, which implements the dirty images as specified in \citet{Briggs_1995}.
}
\label{fig:visibilities}
\end{figure*}

The ``maximum likelihood'' part of RML refers to finding a set of visibilities that maximizes the likelihood function
\begin{equation} \label{eqn:basic_likelihood}
    p\left(\mathbf{D} \mid \mathbf{I}\right).
\end{equation}
The likelihood function expresses the likelihood of measuring a set of visibility data $\mathbf{D}$, given a model image $\mathbf{I}$. The model image can be parameterized in a number of ways. It is possible to proceed with only a handful of parameters to describe the model, for example, a parametric model for a protoplanetary disk might be defined as a set of annular rings, each further specified by their radius and intensity \citep[e.g.][]{Zhang_2016, Guzman_2018}. However, we may not know enough about the source to make such model choices; in this case, a non-parametric approach can offer more flexibility during the imaging process. 

Consider an image with $N \times N$ pixels. Each pixel has some intensity $I_{i}$ such that the image is described by a set of pixel intensities,
\begin{equation}
    \mathbf{I} = \left\{ I_{1}, I_{2}, \ldots, I_{N^{2}}\right\}.
\end{equation}
In this case, the set of predicted model visibilities $\mathbf{V}$ are deterministically calculated by using the Fourier transform of the image sampled at the range of baselines corresponding to the visibility data, such that $\mathbf{V} \overset{\mathcal{F}}{\Longleftrightarrow} \mathbf{I}$. A non-parametric model introduces a great deal of flexibility into the imaging process, which can be harnessed to significantly improve image fidelity compared to a parametric fit \citep[e.g.][]{Jennings_2020}.

We calculate the log likelihood for the sake of computational efficiency. Assuming that the model parameterization (e.g. number of pixels) will remain fixed, and the noise is uncorrelated across baselines and follows a normal distribution with standard deviation $\sigma$, the natural logarithm of the likelihood function is
\begin{equation}
    \ln p\left(\mathbf{D} \mid \mathbf{I} \right) = -N_{\rm{D}} \ln(\sqrt{2\pi}\sigma)-\frac{1}{2} \sum_{i}^{N_{\rm{D}}} \bigg| \frac{\rm{D}_{i}-\rm{V}_{i}(\mathbf{I})}{\sigma_{i}}\bigg|^{2}.
\end{equation}
Here, $N_{D}$ is the number of complex visibilities in the dataset, $\rm{D}_{i}$ is a measured complex visibility at a $(u_{i},v_{i})$ point, and $\rm{V}_{i}$ is the predicted value of the model visibilities for the same $(u_{i},v_{i})$ generated from the model image. Except for the factor of 1/2, the rightmost term above is simply the $\chi^{2}$ statistic,
\begin{equation} \label{eqn:chi2}
    \chi^{2}(\mathbf{D} \mid \mathbf{I})=\sum_{i}^{N_{\rm{D}}} \bigg| \frac{\rm{D}_{i}-\rm{V}_{i}(\mathbf{I})}{\sigma_{i}}\bigg| ^{2},
\end{equation}
and the log likelihood can be expressed as
\begin{equation} \label{eqn:loglikechi}
    \ln p\left(\mathbf{D} \mid \mathbf{I} \right) = -\frac{1}{2} \chi^{2}(\mathbf{D} \mid \mathbf{I}) + C.
\end{equation}
We can now see that in order to maximize the log likelihood, it is necessary to minimize $\chi^{2}(\mathbf{D} \mid \mathbf{I})$. Rather than maximizing the log likelihood, however, in computing it is more common to minimize the negative log likelihood, given by
\begin{equation}
    L_{\mathrm{nll}}(\mathbf{I}) = -\ln p\left(\mathbf{D} \mid \mathbf{I} \right) = \frac{1}{2} \chi^{2}(\mathbf{D} \mid \mathbf{I}).
\end{equation}

In the machine learning community, it is common to focus on the optimization of some metric that can be described by a \textit{loss function} \citep[e.g.][]{Bishop_2006, Hastie_2009, Murphy_2012, Deisenroth_2020}. A loss function is some function which, when minimized, yields optimal parameter values; here we adopt the use of a loss function as the primary quantity to be minimized. It is well established that well calibrated data has Gaussian uncertainties, thus, we adopt the negative log likelihood as the first term in our loss function.

Though the negative log likelihood can function independently as a loss function, it provides no direct constraints on the image, yielding an unregularized fit. In radio interferometry, minimizing the negative log likelihood of the data alone often results in an undesirable image. This is due to the incomplete sampling of the visibility function at certain spatial frequencies; if the visibility function has significant power in $(u,v)$ space that is unsampled or only sparsely sampled, a loss function with no regularization is not particularly useful because there exist many images with the same minimum loss value. As a result, the (dirty) image product is unlikely to be the best representation of the true sky brightness distribution.

Figure \ref{fig:visibilities} shows the $(u,v)$ sampling of the HD~143006 dataset alongside the dirty images made by gridding visibilities with both uniform and Briggs weighting. We use \texttt{MPoL} to generate the dirty images, which implements the same dirty imaging equations as \texttt{CASA}. Uniform weighting yields constant weights within a grid cell, and usually results in an image with high resolution at the cost of sensitivity. Briggs weighting has an adjustable robust parameter which determines the balance between resolution and sensitivity, making it a popular choice for making a visually pleasing dirty image \citep{Briggs_1995}.

The visibility function likely has power at some of the unsampled spatial frequencies. While setting these unsampled but presumably non-zero visibilities to zero is a conventional and conservative imaging procedure, the resulting dirty images contain artifacts such as blotchy emission or a noisy background. One can mitigate the effects of incomplete visibility sampling by regularizing the loss function. Regularizers have different functional forms that can be calculated from the image itself ($\mathbf{I}$) or from quantities derived from the image (e.g. $\mathbf{V}$). For instance, additional terms emphasizing smoothness in intensity between adjacent pixels can be added to the loss function to directly regularize the image. The inclusion of regularizers can greatly influence the visibility function at spatial frequencies not sampled by the interferometer, reducing the number of images that could correspond to the set of observed visibilities and thus lessening the inherently ill-conditioned missing data problem posed by the interferometer. For example, a loss function that includes regularization could be of the form
\begin{equation}
    L(\mathbf{I}) = L_{\mathrm{nll}}(\mathbf{I}) + \lambda_{\mathrm{A}} L_{\mathrm{A}}(\mathbf{I}) + \lambda_{\mathrm{B}} L_{\mathrm{B}}(\mathbf{I}) + \ldots,
\end{equation}
with various loss function formulations suiting specific datasets and science goals. Each $\lambda$ coefficient allows the strength of each regularizer to be tuned. Tuning this parameter is important in order to prevent over-regularizing the model, as a poorly-weighted regularizing term will result in an image that is either not sufficiently different from the dirty image or an image that matches the observed data but imposes an overly strong prior. In a Bayesian framework, these regularizing terms would be akin to prior probability distributions imposed on various parameters of the model, as they impose some existing knowledge or expectation about the source on the model \citep{Sivia_2006}.

\subsection{Regularizers} \label{sec:methods:regularizers}

Implementing regularizers in the imaging process effectively allows us to make assumptions about unsampled and noisily sampled frequencies based on our prior knowledge of the source, in many cases changing the loss function space to become convex and have one clear minimum corresponding to a specific image rather than many minima (and thus many images) that perfectly fit the sampled data. In practice, one may need to use a combination of several regularizers to obtain an image that best represents the true sky brightness.

Regularizers vary in their implementations and their potential effects on the image. Some regularizers can be imposed by construction. For example, certain parameterizations of $\mathbf{I}$ may disallow negative surface brightness values. Other regularizers can be imposed via loss terms, computed directly as a function of the image pixels themselves or via some additional property derived from the image (e.g. the power spectrum). These additional loss terms will require their own strength prefactors $(\lambda_i)$, which can be adjusted to balance the relative impact of each regularizer. Here we discuss the functional form and motivation for each regularizer we tested, selected based on their well-known nature and ability to place sensible constraints on astrophysical images.

\subsubsection{Image Positivity}
The true surface brightness distribution of any astrophysical source will be strictly greater than or equal to zero intensity. This constraint is frequently violated by CLEAN-based imaging procedures, with many synthesized images containing negative pixels in noisy background regions. The physical constraint on image positivity can be naturally incorporated into an RML imaging framework via construction of the image parameterization $\mathbf{I}$.

Rather than directly parameterizing $\mathbf{I}$ using the set of pixel values $\{I_1, I_2, \ldots, I_{N^2} \}$, instead, we parameterize the pixel values using variables $\{Z_1, Z_2, \ldots Z_{N^2}\}$ which are then mapped $Z_i \to I_i$ using a function with a strictly positive range. We chose the Softplus function where $I_i$ is defined by
\begin{equation}
    I_i = f_{\rm{Softplus}}(Z_i)= \log (1+\exp (Z_i)).
\end{equation}
The Softplus function maps negative input values to small but positive non-zero output while leaving positive input values largely unchanged. $I_i = \exp(Z_i)$ is another potential mapping function, however, we found the Softplus function hastened model optimization.

\subsubsection{Maximum Entropy}

Maximum entropy is one of the best-established regularizers for radio interferometric imaging, and has been shown to deliver images with better spatial resolution than the CLEAN algorithm \citep{Cornwell_1985,Narayan_1986}. Maximum entropy regularization aims to find an image that 1) is consistent with all testable information (here, the visibilities sampled by the interferometer) and 2) is maximally non-committal to untestable parameter space \citep{Ables_1974, Sivia_2006}.

Several different functional forms of the maximum entropy regularizers have historically been used, usually similar to either $\log I$ or $-I \log I$ (where the base of the logarithm could be any value, including $e$). The latter is similar in form to statistical mechanics equations of entropy, but repurposed for information entropy \citep{Shannon_1948}. We follow the definition in \citet{EHTCollab_2019} and define maximum entropy loss as
\begin{equation}
    L_{\rm{ent}}=\frac{1}{\zeta} \sum_{i} I_{i} \ln \frac{I_{i}}{p_{i}},
\end{equation}
where $\zeta$ is a normalization factor and $p_{i}$ is a reference pixel value against which other pixels are compared. In this work we used $\zeta = \sum_{i} I_{i}$ The reference pixel values could be as simple as a ``blank'' image of uniform intensity \citep[e.g.][]{Carcamo_2018}, or they could take additional knowledge about the source into account. For example, \citet{EHTCollab_2019} used circular Gaussian images for the sets of $p_{i}$.

Maximum entropy regularization inherently promotes image positivity because of the logarithm built into the functional form of the regularizer; only positive non-zero values $I_{i}$ result in a real and defined $\ln I_{i}$ \citep{Narayan_1986, Hoegbom_1979}. In addition, maximum entropy generally encourages uniform intensities in the image and in the errors, making it a useful regularizer for identifying the presence of features in the image \citep{Hoegbom_1979, Gull_1978}.

Maximum entropy regularization also introduces the potential to achieve some degree of superresolution in the RML image. Superresolution refers to an image that has achieved marked improvement in quality compared to another resolution standard, such as a Gaussian fit to the main lobe of the dirty beam (which is usually but not necessarily the CLEAN beam). The potential for superresolution exists in maximum entropy regularization because the features of the chosen entropy function (e.g. concavity, change in slope) result in an image with sharpened peaks and flattened baseline oscillations \citep{Narayan_1986}. Sharper peaks correspond to resolving features at finer spatial scales, yielding a superresolved image. Flatter baseline oscillations dampen the blotchy imaging artifacts that stem from incomplete sampling of spatial frequencies, such as those seen in Figure \ref{fig:visibilities}.

\subsubsection{Sparsity}

Sparsity regularization uses the $L_{1}$ norm to promote an image that is a sparse collection of non-zero pixels. Derived from the least absolute shrinkage and selection operator \citep[lasso, see ][]{Tibshirani_1996}, sparsity is a pixel-based regularizer that has successfully been applied to radio interferometric imaging to achieve high-resolution images around black holes and protoplanetary disks \citep[e.g][]{Honma_2014, Akiyama_2017b, Kuramochi_2018, EHTCollab_2019, Yamaguchi_2020}. 

We formulate the sparsity loss as
\begin{equation}
    L_{\rm{sparse}}=\sum_{i}\left|I_{i}\right|.
\end{equation}
Sparsity regularization reduces the amplitudes of unneeded pixels (i.e., promoting an image that is a sparse collection of non-zero pixels), making it a useful regularizer when the true sky brightness distribution of a source is likely to be sparse.

The sparsity regularizer does not use any information on the contiguity of blank regions, therefore including a sparsity term will not necessarily favor adjacent bright pixels that would often be expected in a resolved source. However, even if the source is unlikely to be sparse in the image domain (e.g. extended sources like galaxies), sparse regularization has previously been shown to successfully reconstruct these images if the regularization is applied in some other domain like wavelet coefficients \citep{Li_2011, Carrillo_2012, Carrillo_2014}. 

\subsubsection{Total Variation}

Total variation (TV) regularization applies the $L_{1}$ norm to the gradient image, that is, the changes in adjacent pixel intensities in the image.  As a result, TV regularization promotes images with sharp edges at areas with significant changes in intensity and relatively smooth areas in-between, exhibiting sparsity in the gradient image. In other words, the TV regularizer is an edge-preserving noise filter. TV regularization has been used with success on its own and in combination with other regularizers for astronomical interferometric imaging \citep[e.g.][]{Wiaux_2010,Akiyama_2017a,Akiyama_2017b}.

Following \citet{Rudin_1992}, we define the TV loss as
\begin{equation}
    L_{\rm{TV}}=\sum_{l, m} \sqrt{\left(I_{l+1, m}-I_{l, m}\right)^{2}+\left(I_{l, m+1}-I_{l, m}\right)^{2}+\epsilon}.
\end{equation}
The image has dimensions where $l$ corresponds to right ascension and $m$ corresponds to declination. The $\epsilon$ term is an optional softening parameter which determines how pixel-to-pixel variations within the image slice will be penalized. If adjacent pixels vary more than $\epsilon$ the total loss will greatly increase, so TV regularization favors minimal variation between adjacent pixels.

\subsubsection{Total Squared Variation}

The total squared variation (TSV) regularizer is a variant of the TV regularizer, still summing the brightness differences between adjacent pixels. However, by not taking the square root of the differences, the TSV prior results in images with smoother edges \citep{Kuramochi_2018}. The TSV regularizer,
\begin{equation}
L_{\rm{TSV}}=\sum_{l, m}\left(I_{l+1, m}-I_{l, m}\right)^{2}+\left(I_{l, m+1}-I_{l, m}\right)^{2},
\end{equation}
is functionally similar to the TV prior, except the expression inside of the summation has been squared and we no longer include a softening parameter.

\subsection{Minimizing the Loss Function}

\begin{figure*}
\centering
\includegraphics[width=0.9\linewidth]{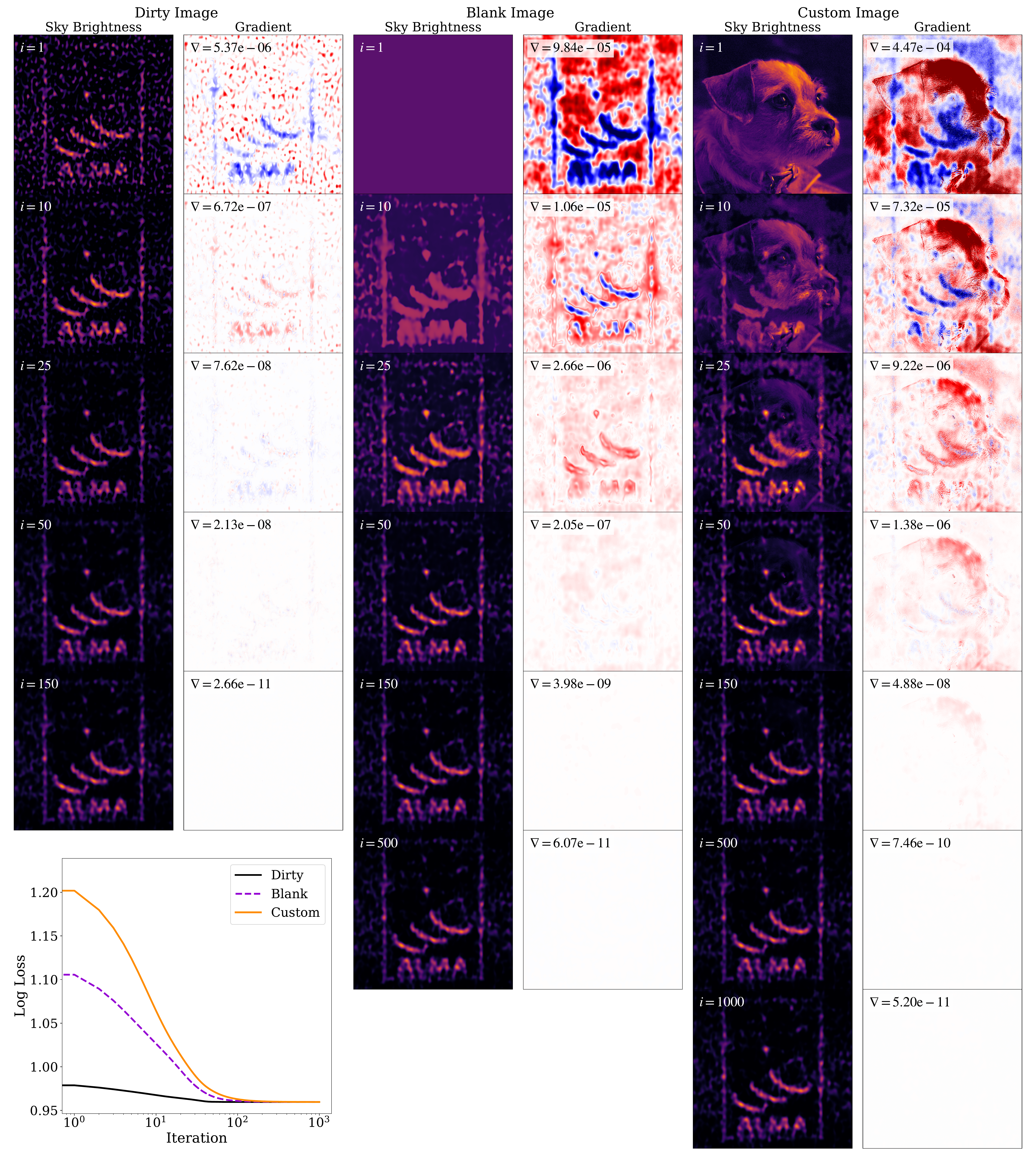} 
\caption{Visualizing the regularization process using gradient descent optimization to minimize the loss function for different initial pixel values.
A combination of entropy ($\lambda=0.5$), sparsity ($\lambda=5\times 10^{-6}$), and total squared variation ($\lambda=10^{-4}$) regularizing terms were used on a mock dataset made from the ALMA logo with added noise. Each row shows the state of the model and gradient image at a different number of iterations during optimization. The leftmost columns show the sky brightness and gradient images of a model initialized with the dirty image, the middle columns show the same for a model initialized with a blank image, and the rightmost columns show the same for a model initialized with a custom image (in this case, an image of a dog). From left to right, this can be read as the best to worst guess for the true sky brightness distribution. In each gradient image, positive (negative) values are shown in red (blue), and $\nabla$ denotes the magnitude of the gradient vector (i.e. the gradient value of each pixel added in quadrature). All sky brightness and gradient images are plotted on the same color scale. A good set of initial pixel values like the dirty image quickly converges to the final image, while a poor guess is more computationally expensive but achieves the same result.
}
\label{fig:init_base}
\end{figure*}

Minimizing the loss function maximizes the likelihood function, giving a ``best fit'' image that can change based on what kind of regularization is implemented. There are a variety of optimization methods that can be used for this minimization problem, such as those that require computing first- or second-order derivatives (e.g. gradient descent algorithms, Newton's method) or those that attempt to minimize a function without computing gradients. We use gradient descent methods, which are iterative processes with several components. First, the gradient of the loss function is computed with respect to model parameters,
\begin{equation}
    \nabla L(\mathbf{I}) = \left\{ \frac{\partial{L(I_{1})}}{\partial{I_{1}}}, \frac{\partial{L(I_{2})}}{\partial{I_{2}}}, \cdots, \frac{\partial{L(I_{N^2})}}{\partial{I_{N^2}}} \right\}.
\end{equation}
Here, the set of model parameters is equivalent to the set of pixel intensities. In order to begin optimization, it is necessary to select an initial set of pixel intensities to be evaluated against the loss function for the first iteration. The simplest starting point is a constant value image. However, a faster alternative is to initialize the model with some approximation of the true sky brightness. For most ALMA datasets, the dirty image itself is already a decent approximation of the sky brightness distribution, enabling the optimization process to converge in fewer iterations than if the initial state of the parameters had been uniform (or in any other configuration that is unlikely to represent the true sky brightness, as shown in the last two columns of Figure \ref{fig:init_base}). If the loss function is convex (i.e. has only a single global minimum), the model will converge to the same result regardless of the initial state of the parameters. The loss surface is convex for the regularizing terms presented here \citep[e.g. see][]{Akiyama_2017b, Chael_2018, Yamaguchi_2020}, so a poor choice of initial pixel intensities comes only at the cost of requiring more iterations to converge on a minimum loss value.

Figure \ref{fig:init_base} shows how different sets of initial pixel intensities impact the speed of convergence while regularizing a sky brightness projection of the ALMA logo with added noise (described in Section \ref{sec:data}). We apply entropy, sparsity, and total squared variation regularizers to the loss function. The dirty image converges first, the blank image second, and the custom image last. We use a custom image of a dog, intentionally selecting a set of pixel intensities with no similarity to the true image. Though the custom set of initial pixel intensities takes significantly longer to converge, it ultimately does converge on the same result as the initial dirty and uniform images, showing that the final result is not sensitive to the initial state of the model. Figure \ref{fig:init_base} also shows how the model image is updated during optimization: after each iteration, the gradient of the image is added to the model parameters, creating a new model image. This process repeats until the loss function converges on a minimum, and the gradient is zero or approximately zero.

One important consideration with the gradient descent method is step size, also called the learning rate. Steps that are too large could overshoot the minimum, causing the algorithm to diverge. The smaller the step size, the more iterations will be required for the loss function to converge on a final value, meaning that steps that are too small can quickly become too computationally expensive to reach the minimum (Ch. 7.1, \citealt{Deisenroth_2020}; Ch. 8.4, \citealt{Murphy_2022}). It is essential to check that the optimization algorithm has converged; an image that has not been fully optimized can be misleading because it is not actually the maximum likelihood solution. For example, in Figure \ref{fig:init_base} all of the model images at 50 and 150 iterations look quite similar. However, the magnitude of the gradient image (defined as the gradient value of each pixel added in quadrature) is reduced by several orders of magnitude at 150 iterations. We can also see that the loss function has not yet been minimized at 50 iterations, especially when the model was initialized with a blank or custom image. Though it may be tempting to run fewer iterations in the interest of computational speed, it is essential to use enough iterations so that the loss function fully converges on a solution.

\subsection{Cross-Validation} \label{sec:methods:cv}

\begin{figure}[h]
\centering
\includegraphics[width=0.75\linewidth]{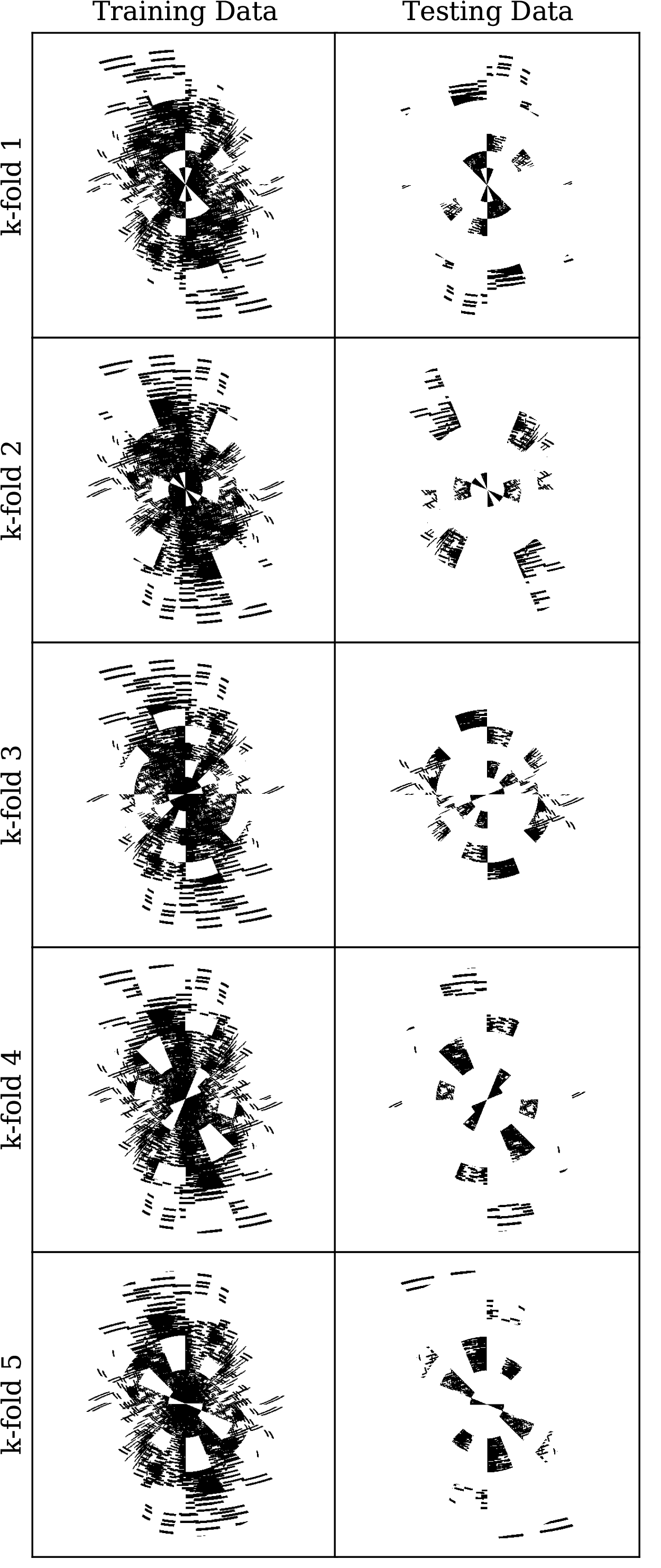}
\caption{Visualizing the data partitioning for K-fold CV ($K=5$) of the HD~143006 dataset using the dartboard scheme with 12 log-linearly spaced radial bins and 8 equal-sized wedges from 0 to $\pi$. The left column shows 4 subsets combined and used to fit the model, while the right column shows the withheld subset of data used for validating the model. Each of the $K$ rows shows a different subset of data used as the testing data.}
\label{fig:kfold_mask}
\end{figure}

Regularizers can be tuned by trial and error, testing new $\lambda$ values until a seemingly reasonable value is found. This method has historically been used with success \citep[e.g.][]{Casassus_2006}, however, modern computational resources enable a more systematic way of determining $\lambda$ prefactor values and optimally tuning regularizers. One way to determine whether the regularization (whether it be the strength of the $\lambda$ prefactors or the functional form of the regularizer itself) is appropriately tuned is by using cross-validation (CV). CV aims to find optimal parameter values by determining how consistently the model performs given variations in the data set, working on the concepts of \textit{training data} and \textit{testing data} (Ch. 7.10, \citealt{Hastie_2009}; Ch. 8, \citealt{Deisenroth_2020}).

Training data is used to find the model which minimizes the specified loss function (including regularizers) and yields the best-fit image. Testing data is used for comparison against the model optimized with the training data. If some range of spatial frequencies is not covered by the training data, but is covered by the testing data, then comparing the trained model to the testing data effectively measures the predictive power of the model with respect to that range of spatial frequencies. In other words, testing data allows us to see how well the model predicts new data.

In principle, one would like to have a large enough pool of data such that partitioning it into a training set and a testing set would not compromise the utility of either subset. When dealing with costly observational data, however, using enough data to train the model typically leaves only a small amount for testing, resulting in a noisy estimate of the predictive performance of the model (Ch. 1.3, \citealt{Bishop_2006}). CV partially circumvents this limitation by partitioning all the measured visibility data $\mathbf{D}$ into subsets such that ${\mathbf{D}_{i}}(\{u,v\}) \subset \mathbf{D}$, fitting the model on one or more subsets, and testing the model on the remaining subsets. One popular method is K-fold CV, which performs this process in multiple rounds and rotates which subsets are used for testing in each round \citep[e.g.][]{Akiyama_2017b, Akiyama_2017a, Yamaguchi_2020}. In this case, data are partitioned into $K$ subsets. 
\begin{equation}
    \mathbf{D} \left\{
    \begin{aligned}
        &{\mathbf{D}_{1}}(\{u,v\}) \\
        &{\mathbf{D}_{2}}(\{u,v\}) \\
        &\vdots \\
        &{\mathbf{D}_{K}}(\{u,v\})
    \end{aligned}
    \right.
\end{equation}
After partitioning, $K-1$ subsets are combined to form the training data $\mathbf{D}_{\rm train}$ and the remaining subset $\mathbf{D}_{\rm test}$ is used for testing. Using only $\mathbf{D}_{\rm train}$, a model image $\mathbf{I}_{\rm train}$ is generated. The full visibility function $\mathbf{V}_{\rm train}$ is obtained from $\mathbf{I}_{\rm train}$ using the fast Fourier transform. Finally, $\mathbf{V}_{\rm train}$ and the withheld test set $\mathbf{D}_{\rm test}$ are compared within the same $(u,v)$ space originally sampled by $\mathbf{D}_{\rm test}$. Applying Equation \ref{eqn:loglikechi}, we obtain
\begin{equation}
    p\left(\mathbf{D}_{\rm test} \mid \mathbf{I}_{\rm train} \right) \propto \exp \left[-\frac{1}{2} \chi^{2}(\mathbf{D}_{\rm test} \mid \mathbf{I}_{\rm train})\right],
\end{equation}
which indicates that a smaller $\chi^{2}$ value corresponds to a better match between the trained visibilities and the testing data. In other words, the lower the $\chi^{2}$ value, the higher the probability of $\mathbf{V}_{\rm train}$ accurately modeling visibilities not included in the original data set.

This process is repeated $K$ times such that each subset functions as the testing data exactly once. We obtain a final CV score by summing the $\chi^{2}$ values calculated for each of the $K$ CV rounds,
\begin{equation}
    \rm{CV} = \sum^{K}_{k} \chi^{2}(\mathbf{D}_{k} \mid \mathbf{I}_{\rm train}),
\end{equation}
where $\mathbf{D}_{k}$ is the test dataset, and $\mathbf{I}_{\rm train}$ is the model image that minimizes the loss function for the training data $\mathbf{D} - \mathbf{D}_{k}$. A low CV score indicates that the model (consisting of the choices of image parameterization, regularizers, and regularizer strengths), when trained on the training data, does a good job at predicting the withheld training data. If the regularizers and their strengths are poorly chosen, however, at least two failure modes arise. In the first, the model may simply fail to fit the training data adequately. This can happen if the model is over-regularized (not sufficiently flexible). When this happens, it is not surprising that the model also fails to predict the withheld test data accurately. The second failure mode arises when the model fits the training data accurately but fails to predict the withheld test data. This can happen if the model is under-regularized. In this situation, the model would be said to be over-fit.

Visibility datasets acquired by ALMA have many unique characteristics, such as their number of samples, variable density of $(u,v)$ sampling, and varying signal-to-noise ratio, when compared to simpler datasets (e.g., data points in a polynomial regression). This presents many opportunities and challenges for how to partition data for K-fold CV. We explored CV using two methods of partitioning, which we dub ``random cell'' and ``dartboard.'' Random cell partitioning utilizes $K$ subsets that are composed of randomly-selected visibility grid cells. Grid cells are randomly drawn without replacement so that each grid cell is only used in a single subset. The one exception is that, for numerical stability, we ensure that $(u,v)$ cells with the highest 1\% of gridded weight values are included in each $K$ subset. These cells are usually those at the shortest baselines and most informative about the total flux of the source. Dartboard partitioning uses polar grid lines to create a new layer of azimuthal and radial bins, each of which contains many visibility grid cells. Each of the $K$ subsets consists of randomly-drawn dartboard cells without replacement. Figure \ref{fig:kfold_mask} shows an example of dartboard partitioning.

\subsection{The MPoL Package} \label{sec:methods:mpol}
Million Points of Light (\texttt{MPoL})\footnote{\url{https://mpol-dev.github.io/MPoL/}} is an open-source Python package we have designed as a foundation to enable RML imaging for a variety of interferometric workflows. \texttt{MPoL} is built on PyTorch \citep{PyTorch_2019}, an open-source machine learning framework that provides a ``tensor'' array with the ability to calculate gradients using auto-differentiation. Gradient calculations with auto-differentiation enable users to easily and rapidly minimize a loss function with gradient descent methods, as illustrated in Figure \ref{fig:init_base}. 

Even with fast calculation of gradients, minimizing the loss function for RML images with many pixels can quickly become computationally expensive on a CPU. Because computational time scales with image dimensions, parameterizing an image with more pixels results in a slower RML imaging process. This can be further exacerbated for data cubes with both a large number of pixels and many channels. Although RML imaging techniques, in particular maximum entropy, have existed for decades, the required computational resources placed substantial limitations on the sizes of the synthesized images. For this reason, \texttt{MPoL} takes advantage of the power of GPUs, which can greatly reduce computation time compared to CPUs. For a single $1024 \times 1024$ pixel RML image with \texttt{MPoL}, computation time tends to be a few minutes on a CPU and a few seconds on a GPU, though this will vary depending on the number of iterations needed to reach convergence. This is relatively fast either way, especially compared to CLEAN methods which may take days for high resolution ALMA observations that require many CLEAN components to synthesize the image.


\section{Results} \label{sec:results}

\begin{figure*}
\centering
\includegraphics[width=0.95\linewidth]{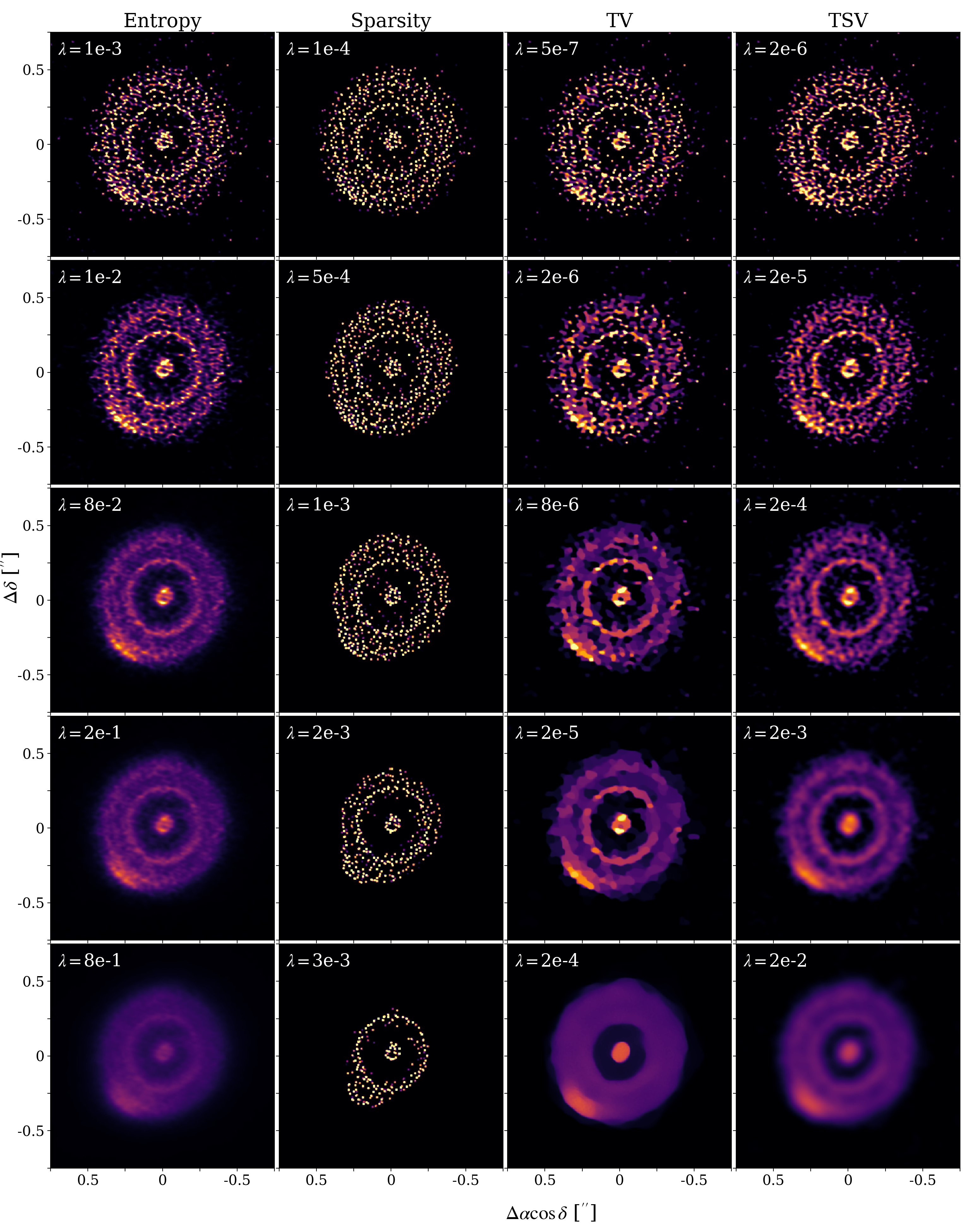}
\caption{The effect of regularizers on images of HD~143006, applied one at a time. From left to right, each column shows entropy, sparsity, total variation, and total squared variation regularizers at varying strengths. The top row uses the least regularization (i.e. a small $\lambda$ prefactor on the regularizing term) and the bottom row shows extremely high regularization. A range of $\lambda$ values were selected in order to show the full range of possible images; near optimal $\lambda$ values can be found using CV methods. All images are displayed on the same color scale.}
\label{fig:image_grid}
\end{figure*}

\begin{figure*}
\centering
\includegraphics[width=0.95\linewidth]{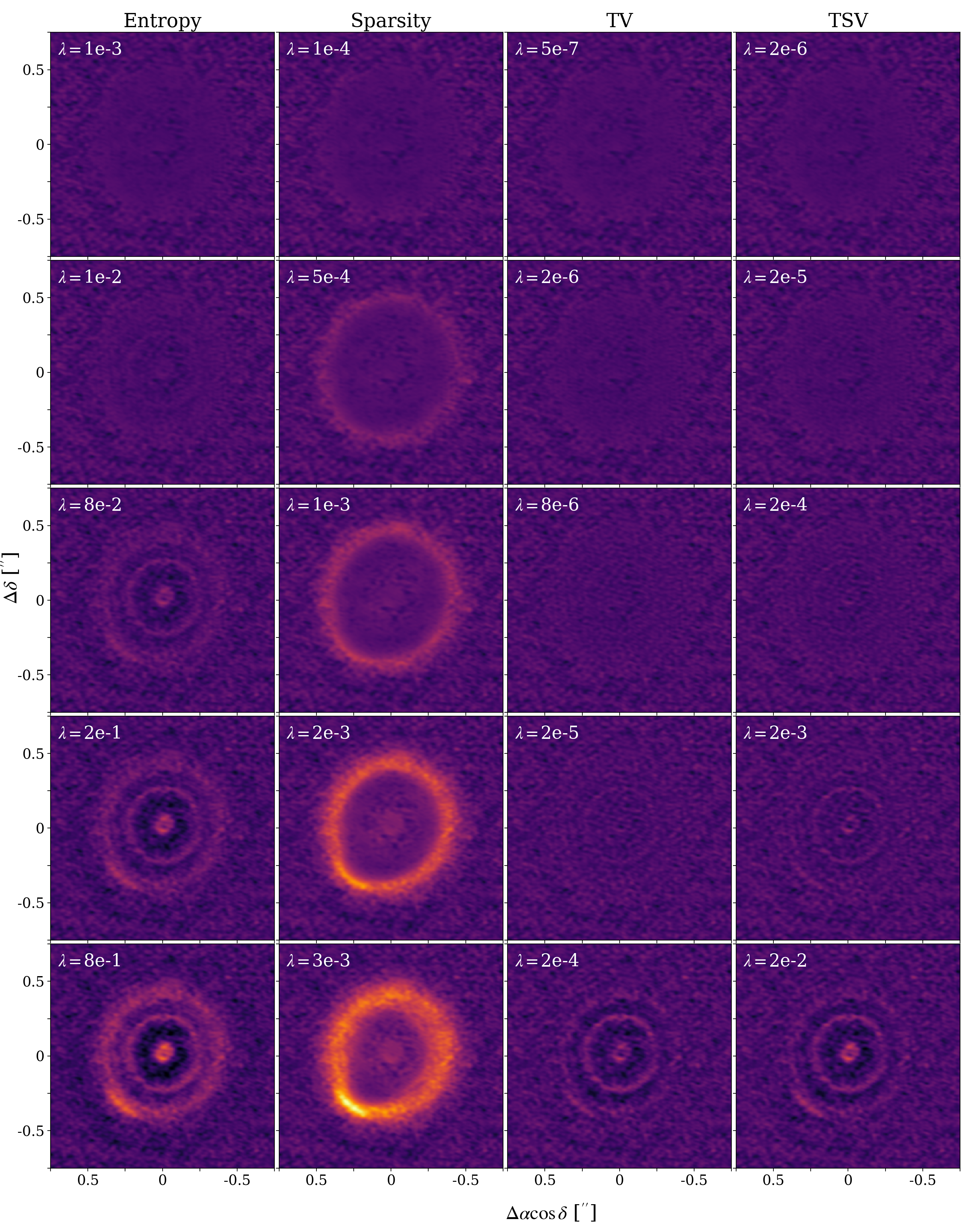}
\caption{Residuals for the images presented in Figure \ref{fig:image_grid}. Regularizer strengths are lowest in the top row, and highest in the bottom row. Images generated with strong regularizing terms show structure in the residual images, indicating that the model is underfitting the data. All images use Briggs weighting (robust$=0.0$). All images are displayed on the same color scale.}
\label{fig:residual_grid}
\end{figure*}

We explored the effects of entropy, sparsity, TV, and TSV regularizers on images produced from both the simulated protoplanetary disk dataset and the real HD~143006 dataset described in Section~\ref{sec:data}. Figure \ref{fig:image_grid} shows the result of each of these regularizers at different strengths (set with the $\lambda$ prefactor on each term). The images shown were generated using an arbitrarily chosen range of $\lambda$ values in order to show the breadth of images possible with different regularizer tunings. Figure \ref{fig:residual_grid} shows the residuals from each optimized image for each of the panels shown in Figure \ref{fig:image_grid} imaged from the residual visibilities using the \texttt{DirtyImager} with Briggs weighting (robust = 0.0). While we found the most success using multiple regularizers in combination with each other, here we qualitatively describe the effect of each regularizer on protoplanetary disk images in isolation. Figures \ref{fig:image_grid} and \ref{fig:residual_grid} show variation across images made from the HD~143006 dataset; we verified that the simulated disk dataset exhibits similar behavior.

\subsection{Entropy Performance}

The first column of Figure \ref{fig:image_grid} shows the effect of different $\lambda$ prefactors for maximum entropy regularization of HD~143006 with a positive, uniform set of reference pixels ($\mathbf{I}=10^{-7} \; \rm{Jy \; arcsec}^{-2}$). Even at high $\lambda$ values, maximum entropy regularization can retain high-resolution features in the image. However, because maximum entropy regularization generally promotes uniformity in the image, the image tends to a model image that appears ``faded'' at excessive values of $\lambda$, making emission appear fainter across the entire source. The bottom-left panel of Figure \ref{fig:image_grid} shows an example of such an image.

The primary indication of over-regularization with maximum entropy is an image that appears faint or slightly blurred compared to images made from different entropy $\lambda$ values, suppressing bright peaks in the image. Another way to check for over-regularization is by examining the residual image. In the bottom-left panel of Figure \ref{fig:residual_grid}, ringed structure is evident in the residual image created by maximum entropy regularization with $\lambda=8\times 10^{-1}$. In some cases, these effects may be mitigated by using a non-uniform set of reference pixels, such as a circular Gaussian \citep[e.g.][]{EHTCollab_2019} or a uniform ring. This should be done with caution, as maximum entropy regularization favors similarity with the reference image, and the reference image may not capture enough characteristics of the true source. This could have unintended consequences such as regularizing out small-scale structures (for example, localized asymmetries) that are present in the data but not in the set of reference pixels.

\subsection{Sparsity Performance}

Sparsity regularization promotes mostly blank images, with only the most impactful pixels having non-zero values. Only a $\lambda$ prefactor determines how strongly sparsity should be imposed during optimization; unlike maximum entropy regularization, no reference image is needed. Column 2 of Figure \ref{fig:image_grid} shows images of HD~143006 with 5 different sparsity $\lambda$ values. A small $\lambda$ can effectively suppress noisy background pixels in the image without changing much, if anything, about the source emission. Sparsity regularization alone does not introduce any ``smoothing'' effects that may be desirable for a resolved source; the image will ultimately be a sparse collection of non-zero pixels, which can make it difficult or impossible to identify small-scale structures within the image. For protoplanetary disk continuum datasets, sparsity regularization is most effectively used in combination with other regularizers.

Over-regularization with sparsity can have a significant negative impact on image fidelity, yielding an image that is not representative of the entire source. Because the sparsity regularizer encourages mostly blank images, one potential drawback is the risk that astrophysically real but faint emission may not appear in the synthesized image. This has the effect of neglecting more diffuse emission in the synthesized image, as diffuse emission lacks bright peaks for the sparsity regularizer to identify. In the case of HD~143006, sparsity regularization with a high $\lambda$ value removed some of the more diffuse emission, and in extreme cases removed some rings entirely. In Figure \ref{fig:image_grid}, the bottom panel of column 2 shows the result of imaging HD~143006 with sparsity over-regularization. Here, the outer ring has been regularized away, but the bright azimuthal asymmetry that normally coincides with the outer ring is still present, completely misrepresenting the morphology of the source.

Over-regularization is very evident in the residual image (column 2, Figure \ref{fig:residual_grid}). Because only the most prominent features remain in the model image, any diffuse or generally lower-intensity emission will instead be apparent in the residuals. In the model image of a resolved source, things to look out for that may indicate sparsity over-regularization include features that appear `incomplete' such as having partial rings, an unexpected bright standalone feature, or unexpectedly sharp changes in intensity.

\subsection{Total Variation Performance}

TV regularization promotes sharp edges between areas of different intensities, with smoothness in areas of similar intensity. Column 3 of Figure \ref{fig:image_grid} shows the effect of different $\lambda$ prefactors for TV regularization of HD~143006. Because TV promotes similarity between adjacent pixels unless there is a large change in intensity (i.e. sparsity in the spatial gradient of the image), TV regularization can result in a model image composed of many nearly uniform cells, each containing several pixels. This can create an optical illusion where the image appears to have larger pixels than the true pixel size (e.g. see the image in the third column and third row of Figure \ref{fig:image_grid}, where $\lambda=8 \times 10^{-6}$).

This effect changes as the $\lambda$ value increases, with model images being reminiscent of a watercolor painting or a photo that has been posterized. Because TV regularization does not favor gradual changes in intensity, instead preferring sharp changes, a smooth change in source intensity is likely to become a set of sharply defined layers in the image. In the case of extreme over-regularization, this can remove most detail from the image, resulting in an image that appears blotchy or smeared. However, in a source like HD~143006 which exhibits ringed emission, the smearing is mostly azimuthal rather than radial, retaining some large scale ring structure while losing or minimizing finer details like gaps or azimuthal asymmetries.

For these reasons, TV regularization may be a poor choice if the source is likely to have small scale features or gradual changes in intensity, as many astronomical sources do. Though over-regularization can be evident due to the presence of structure in residual images (see the bottom panel of column 3, Figure \ref{fig:residual_grid}), it may not be evident from inspection of residuals alone until well beyond the $\lambda$ value at which morphological details are regularized out of the image.

\subsection{Total Squared Variation Performance}

Like TV regularization, TSV regularization promotes sharp edges between areas of different intensities. However, the TSV regularizer is less rigid with this condition, allowing for larger differences between adjacent pixels --- while TV regularization applies sparsity (or the L1 norm) to the gradient of the image, TSV regularization applies the L2 norm to the gradient of the image. This makes TSV a strong performer for sources with clearly defined but not perfectly sharp features, such as ringed emission. The rightmost column of Figure \ref{fig:image_grid} shows images of HD~143006 with 5 different TSV $\lambda$ values. Well-tuned TSV regularization performs comparably to maximum entropy regularization, retaining high-resolution features in the model image.

The primary sign of over-regularization with TSV is a blurred image. If the TSV-regularized image appears to have no sharp features at all, as if it had been put through a low pass filter, it is likely over-regularized. This is also evident in the residual images (rightmost column, Figure \ref{fig:residual_grid}), where sharp structures will appear if they have been regularized out of the sky brightness image.

\subsection{Hyperparameter Tuning}

\begin{figure*}[ht]
\centering
\includegraphics[width=0.8\linewidth]{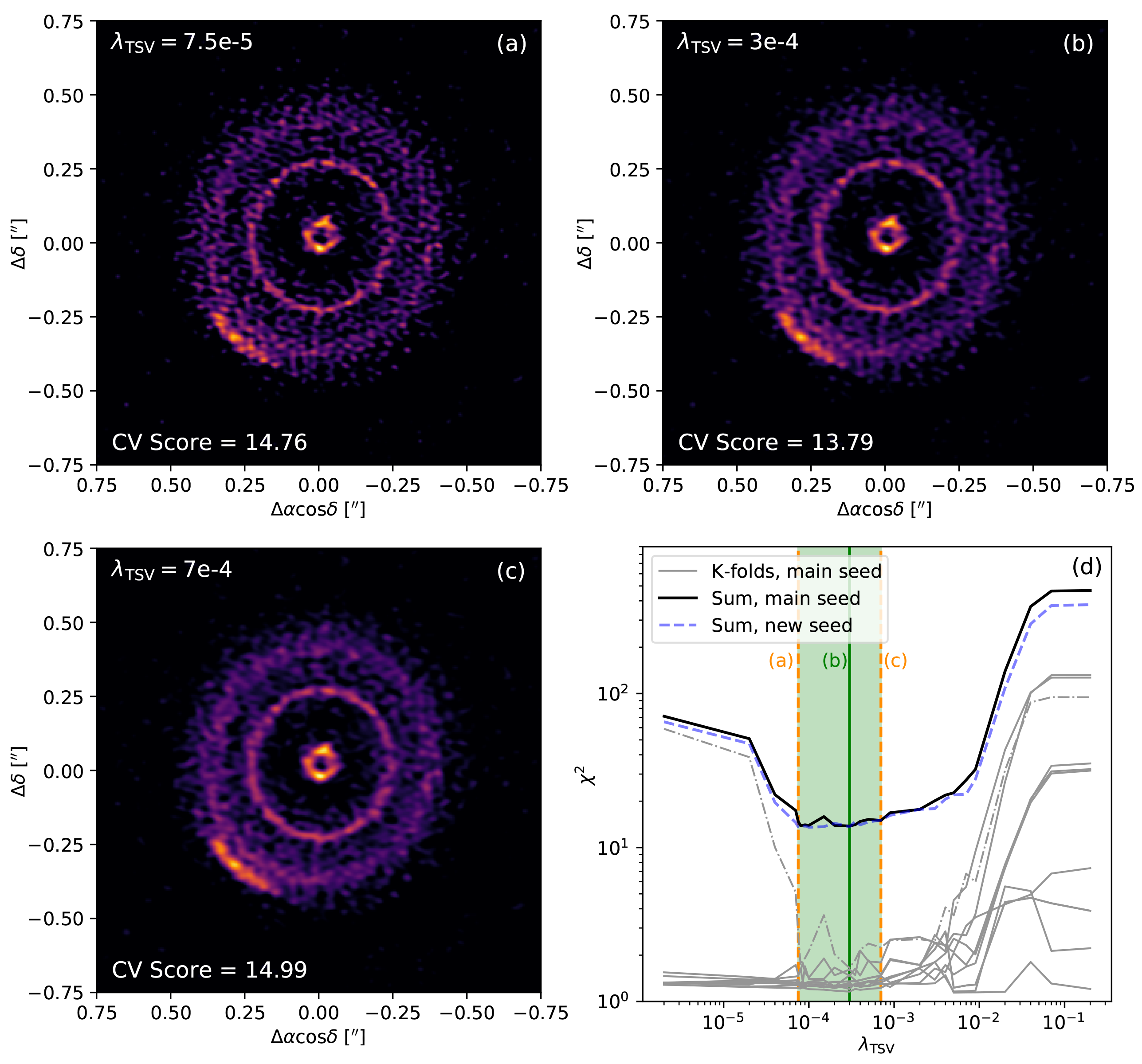}
\caption{An example of hyperparameter tuning with CV scores, using dartboard train/test set partitioning with 30 log-linearly spaced radial bins and 10 equally spaced azimuthal bins from 0 to $\pi$. CV scores are computed by taking the average $\chi^2$ difference in the model visibilities. Panels a, b, and c show images made with different values of $\lambda_{\rm{TSV}}$ and their corresponding CV scores. Panel b shows the image corresponding to the minimum CV score, while panels a and c show images with CV scores within 10\% of the minimum. Panel d shows how the CV score varies with $\lambda_{\rm{TSV}}$. The gray lines show the $\chi^{2}$ contribution from each of the 10 K-folds used during CV. The solid black line shows the sum of these $\chi^{2}$ contributions, which yields the total CV score. The minimum occurs at $\lambda_{\rm{TSV}} = 3 \times 10^{-4}$. The blue dashed line shows the total CV score for a different random seed used when partitioning data for CV, all else held constant. Here, the minimum occurs at $\lambda_{\rm{TSV}} = 10^{-4}$. Because the CV score with respect to $\lambda_{\rm{TSV}}$ generally makes an asymmetrical U shape, there are a range of $\lambda_{\rm{TSV}}$ values that produce comparably low CV scores. All three images presented here fall into this range which spans nearly a full order of magnitude in $\lambda_{\rm{TSV}}$. The K-fold denoted by the dash-dot line represents the K-fold in which the training data did not include the centermost visibilites (but the testing data did), which dominates the total CV score at low $\lambda_{\rm{TSV}}$.
\label{fig:cv_variation}}
\end{figure*}

\begin{figure*}
\centering
\includegraphics[width=\linewidth]{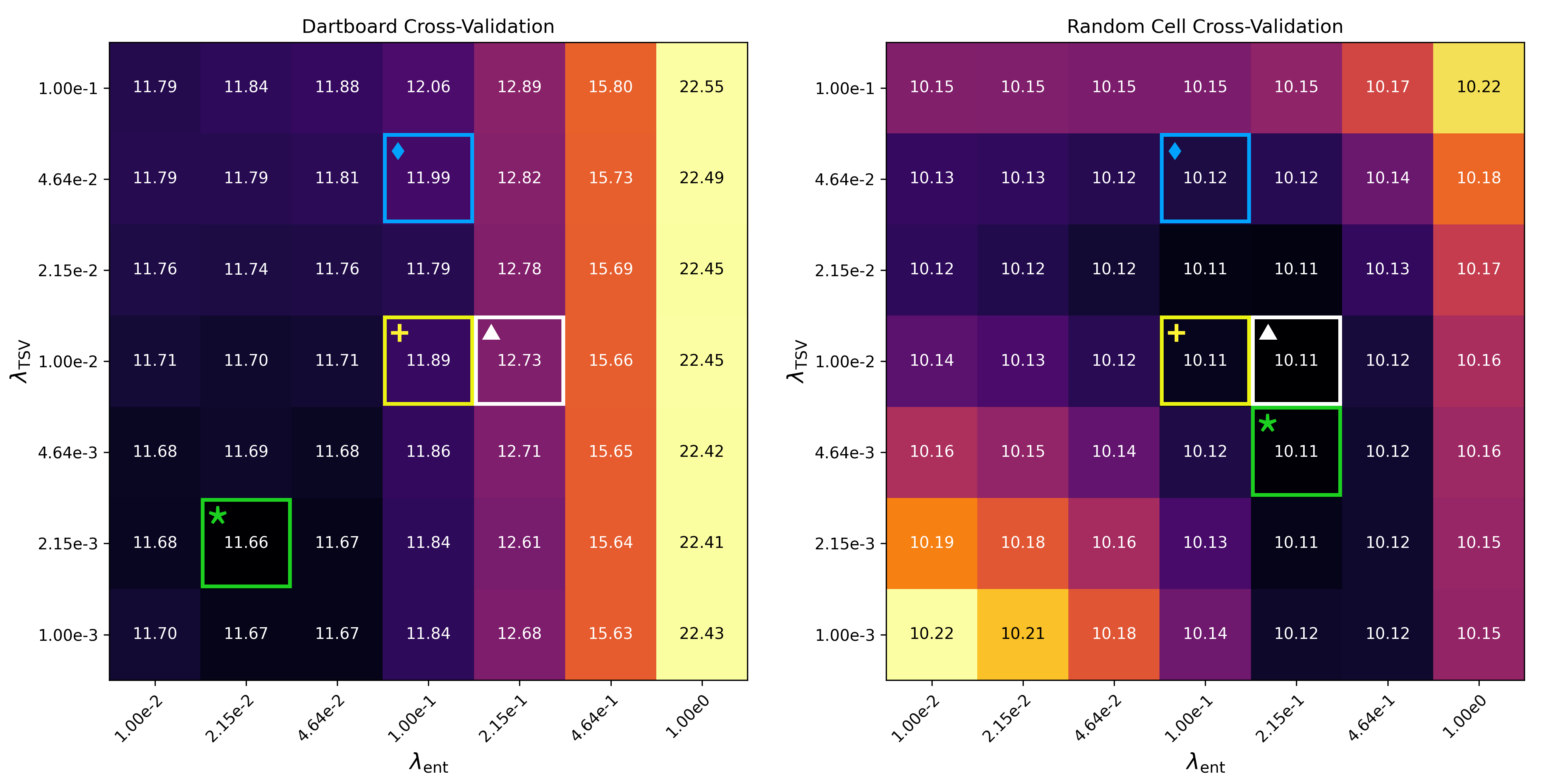}
\caption{CV scores at different hyperparameter setting of $\lambda_{\rm{ent}}$ and $\lambda_{\rm{TSV}}$ for the simulated disk dataset using dartboard (left) and random cell (right) visibility partitioning. CV scores boxed in green ($\star$) are the minimum value found for the specified partitioning method. Scores boxed in yellow (\texttt{+}) correspond to the hyperparameter settings found by adjusting hyperparameters by trial and error until a visually pleasing image is achieved. Scores boxed in blue ($\scriptstyle\blacklozenge$) correspond to the hyperparameter settings which minimized the total pixel difference between the RML model and ground truth, while scores boxed in white ($\blacktriangle$) correspond to the hyperparameter settings which minimized the NRMSE between the RML model and ground truth. Note that hyperparameter settings found for tuning by hand, minimizing the pixel difference, and minimizing the NRMSE are the same in both panels; because they are not CV methods, they do not use any visibility partitioning. 
\label{fig:dartboard_vs_randcell}}
\end{figure*}

\begin{figure*}
\centering
\includegraphics[width=\linewidth]{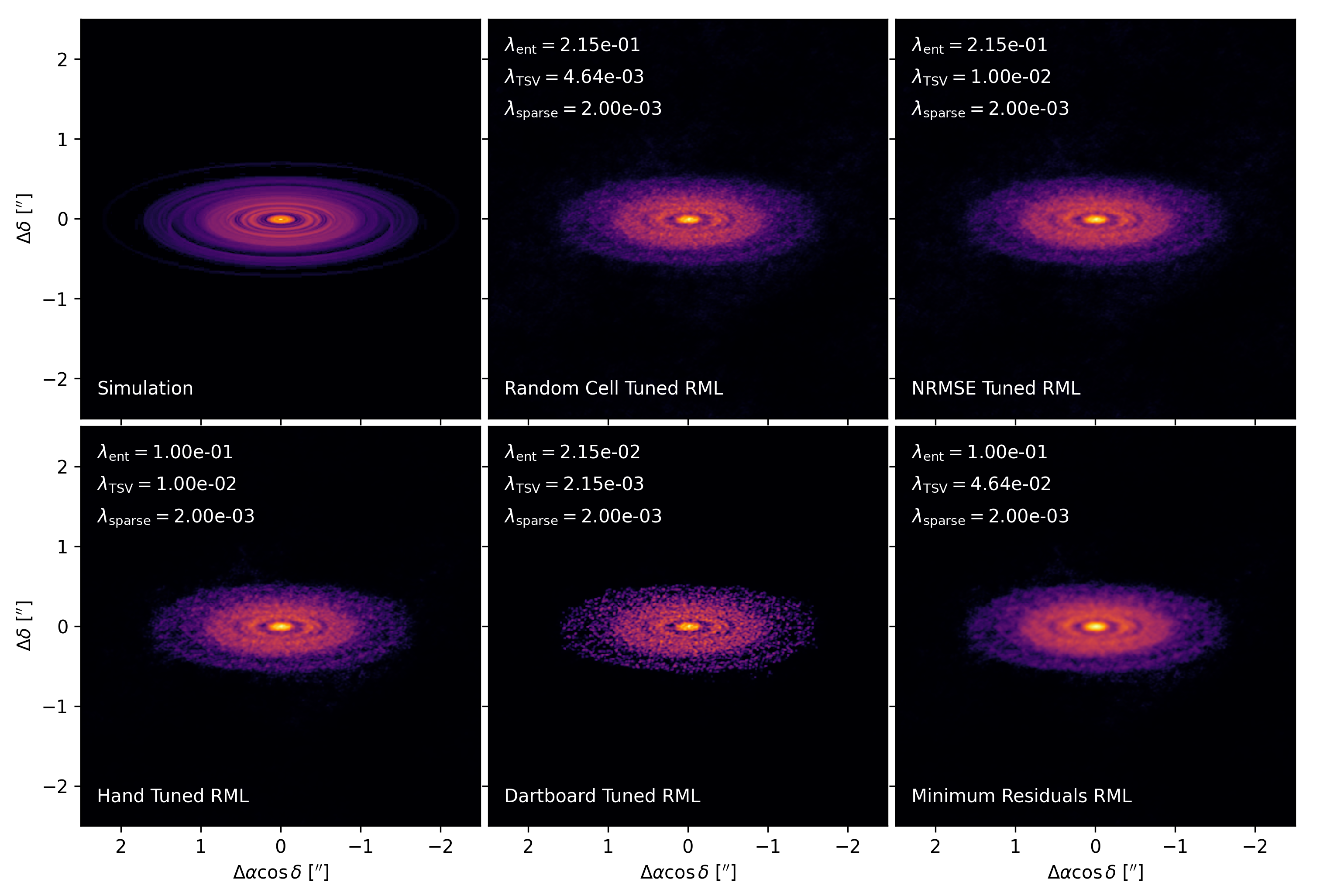}
\caption{The simulated protoplanetary disk image (top left) and five RML images generated using different hyperparameter tuning techniques. The bottom left image was tuned by hand. The top center and bottom center images were tuned by CV using random cell and dartboard visibility partitioning, respectively. The top right image was tuned by minimizing the NRMSE between the RML and true simulated images, and the bottom right image was tuned by minimizing the difference between the RML and true simulated images.}
\label{fig:cv_method_compare}
\end{figure*}

\begin{figure*}
\centering
\includegraphics[width=\linewidth]{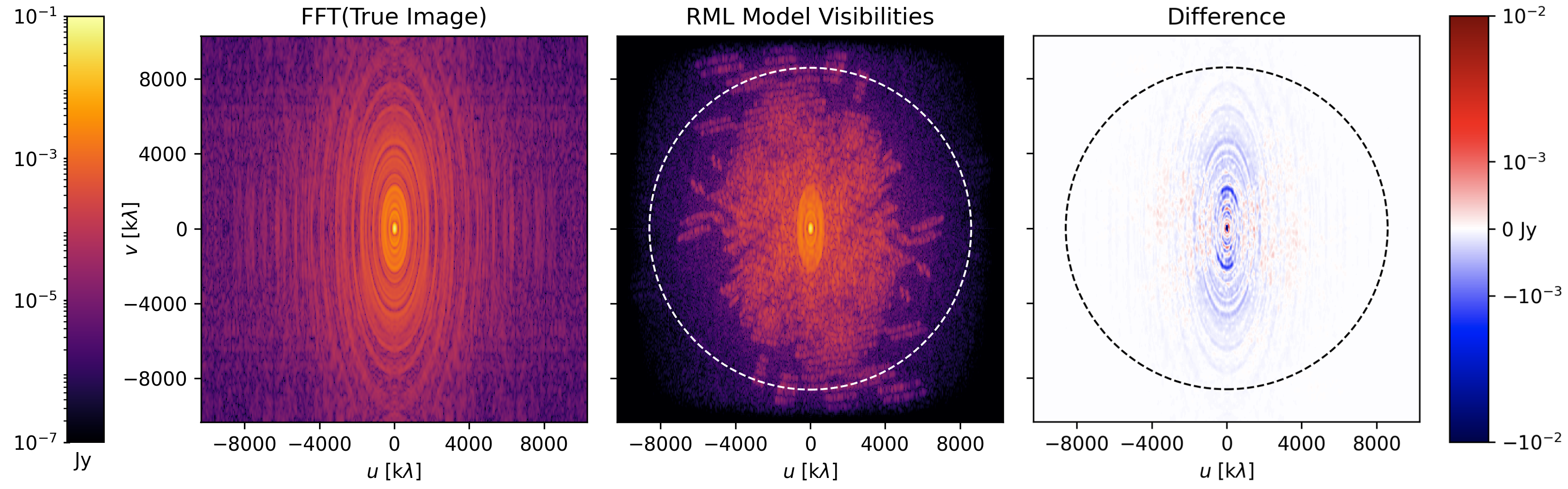}
\caption{A comparison of the true visibility function (left), recovered visibility function using RML (middle) and the residuals (right), shown using the amplitude only. The true visibility function was computed with the FFT of the ground truth image; the RML visibilities are retrieved directly from the forward model. The visibility amplitudes are plotted on the same logarithmic color scale. The right panel shows the difference between the visibility amplitudes of the RML model and the true image, plotted on a symmetrical logarithmic color scale, which switches to a linear scale at absolute values smaller than $2.9 \times 10^{-3}$ Jy (the mean $\sigma$ of non-zero gridded visibility cells with baselines $3000 \;\rm{k}\lambda < q < 5000 \;\rm{k}\lambda$, where each gridded cell is $1623 \;\rm{k}\lambda^{2}$). The circles show spatial frequencies corresponding to a 24 mas beam, which is the approximate resolution of the RML image.}
\label{fig:vis_model}
\end{figure*}

Figure \ref{fig:cv_variation} shows how random variation can impact which hyperparameters minimize the CV score. We demonstrate this with the simple case of tuning a single hyperparameter: imaging the HD~143006 dataset with only TSV regularization and using dartboard visibility partitioning (described in Section \ref{sec:methods:cv}). The CV score with respect to $\lambda_{\rm{TSV}}$ is loosely U-shaped, and there are a range of $\lambda_{\rm{TSV}}$ values that produce comparably low CV scores. We show three images over nearly a full order of magnitude in $\lambda_{\rm{TSV}}$ that correspond to CV scores within 10\% of the minimum CV score. Within a single CV setup, each K-fold also exhibits variation. In particular, the K-fold which lacks the shortest baseline visibilites in the training data tends to dominate the total CV score at low $\lambda_{\rm{TSV}}$. If the random seed for visibility partitioning is changed, this behavior repeats with small variations; the minimum CV score occurs at $\lambda_{\rm{TSV}} = 10^{-4}$ instead of $\lambda_{\rm{TSV}} = 3 \times 10^{-4}$. Thus, rather than requiring strict minimization of the CV score, there commonly exists a range of hyperparameters that result in similar images. This makes tuning with CV methods easier and faster, as one can conduct CV over a fairly coarse grid of hyperparameter values.

We explored the quality of images produced using CV applied to real ALMA data of the disk around HD~143006, but these exercises lacked a comparison around a ``ground truth'' source image. To further evaluate the performance of these two CV schemes, we conducted additional CV analyses using the simulated dataset and its corresponding source image. We find that both random cell and dartboard partitioning as presented in Section~\ref{sec:methods:cv} are viable options for hyperparameter tuning, though random cell may minimize differences between the RML image and ground truth slightly better. A comparison of hyperparameter tuning using both dartboard and random cell visibility partitioning for the simulated disk dataset is shown in Figure \ref{fig:dartboard_vs_randcell}. The left panel shows the CV scores for various $\lambda_{\rm{ent}}$ and $\lambda_{\rm{TSV}}$ for dartboard partitioning, while the right panel shows the same for random cell partitioning. In each panel, the green box highlights the minimum CV score for that particular partitioning scheme, while the yellow boxes show the hyperparameter values selected by tuning by hand (adjusting the hyperparameter values manually until a seemingly reasonable image is achieved).

We find the minimum residual image for each combination of hyperparameters by summing the absolute value of the difference, 
\begin{equation}
    \sum_{i=1}^{N^{2}}\left|I_{i, \rm{true}} - I_{i, \rm{RML}}\right|,
\end{equation}
where $I_{i, \rm{RML}}$ is an RML image pixel and $I_{i, \rm{true}}$ is some corresponding reference image pixel for an image with $N^2$ total pixels. The blue boxes in Figure \ref{fig:dartboard_vs_randcell} show the hyperparameter setting that minimized the difference between the RML and ground truth image (the ``minimum residual'' image). We also compute the normalized root mean squared error (NRMSE),
\begin{equation} \label{eqn:nrmse}
   \rm{NRMSE} = \sqrt{\frac{\sum_{\mathit{i}=1}^{\mathit{N}^{2}}\left|\mathit{I}_{\mathit{i}, \rm{RML}}-\mathit{I}_{\mathit{i}, \rm{true}}\right|^{2}}{\sum_{\mathit{i}=1}^{\mathit{N}^{2}}\left|\mathit{I}_{\mathit{i}, \rm{true}}\right|^{2}}}.
\end{equation}
The hyperparameter setting which minimized the NRMSE between the RML and ground truth image is shown boxed in white.

Each of these tuning methods (random cell and dartboard CV, along with tuning by eye and minimizing quantities derived from the ground truth) yielded similar results. In Figure \ref{fig:cv_method_compare}, we compare the true simulated disk image (top left) with five RML images that correspond to a different hyperparameter tuning method highlighted in Figure \ref{fig:dartboard_vs_randcell}. The bottom left panel shows an image tuned by trial and error, adjusting the hyperparameter values by hand until a seemingly reasonable image is achieved. The top center and bottom center panels show RML images tuned by CV using random cell and dartboard visibility partitioning, respectively. Lastly, the top right panel shows the RML images that minimizes the NRMSE, and the bottom right panel shows the minimum residual RML image.

Both the minimum NRMSE and minimum residual images are obtained by comparing the RML model to the ground truth image in different ways. If we assume that the minimum NRMSE image defines the optimal solution, then random cell CV or simply hand tuning hyperparameters performed best. If we assume that the minimum residual image defines the optimal solution, then hand tuning hyperparameters performed best, closely followed by random cell CV. Despite small variations in the hyperparameter settings, the minimum NRMSE and minimum residual images appear qualitatively similar, though the minimum residual image exhibits a greater degree of smoothing. Conversely, CV with dartboard partitioning resulted in a synthesized image which looks noisier and less smooth than the others. In all cases, however, we found that the RML images successfully recovered the most prominent rings and gaps in the original simulation, but failed to suppress enough noise to recover features at the smallest spatial scales or with low contrast flux variations.

Using the simulated dataset created an opportunity to visualize the performance of the RML model in the visibility domain relative to the ``true'' visibility function. Figure \ref{fig:vis_model} shows how the visibility amplitudes of the simulated protoplanetary disk image (computed by taking the FFT of the ground truth image) compare to the visibility amplitudes of the RML model tuned with random cell CV. The right panel shows the differences between the visibility amplitudes of the RML model and the true image, most of which are on the same scale as the noise of a typical non-zero gridded visibility cell (cell size $= 1623 \;\rm{k}\lambda^{2}$; mean $\sigma = 2.9\times 10^{-3}$ Jy, for baselines $3000 \;\rm{k}\lambda < q < 5000 \;\rm{k}\lambda$). The visibility amplitudes of the RML model only deviate slightly from the visibility amplitudes computed from the ground truth image ($\sim$mJy fluctuations or smaller), demonstrating that RML imaging techniques achieve model visibilities close to corresponding true values.

Even though the visibility amplitudes are generally similar, the baseline distribution of the original data (shown in the left panel of Figure \ref{fig:visibilities}) leaves a distinct imprint on the RML model visibilities; the RML model visibility amplitudes are greatest at $(u,v)$ included in the dataset. The model appears to interpolate well at $(u,v)$ that is already thoroughly sampled by the original data (i.e. $q<4000$), but only partially extrapolates to long baselines. At long baselines, the RML model visibilities have power at $(u,v)$ not represented in the original data, but these regions of the RML model have less power than the true visibility function by at least an order of magnitude. There is also distinct ringed structure in the true visibility function that is largely absent in the RML model visibilities. This structure arises from small flux variations in the visibility function and is only visible on a logarithmic color scale; the noise added to the mock dataset may be one impediment preventing the RML model from reproducing these subtle features.

With any image synthesis method, the optimal image product will depend on the needs of a particular science case. It follows that choices about image validation will also vary depending on which image characteristics are being prioritized (e.g. maximizing sensitivity or achieving the finest possible resolution). Regardless of which characteristics are favored, we find that none of the hyperparameter tuning methods explored here result in RML images that grossly misrepresent the morphology of the source. Combining multiple regularizers is an effective way to achieve a high quality image product, as each regularizer contributes different qualities to the image (e.g. high resolution features from entropy, reduced background noise from sparsity). We recommend using CV methods as a technique to identify hyperparameter settings that improve model image fidelity rather than using an absolute minimum CV score to identify a single ``best'' choice of regularizers and hyperparameter values, as trivial variations in image pixel values appear at hyperparameter settings near the minimum CV score, and the minimum CV score itself exhibits variation stemming from randomness in the visibility partitioning.

\begin{figure*}
\centering
\includegraphics[width=\linewidth]{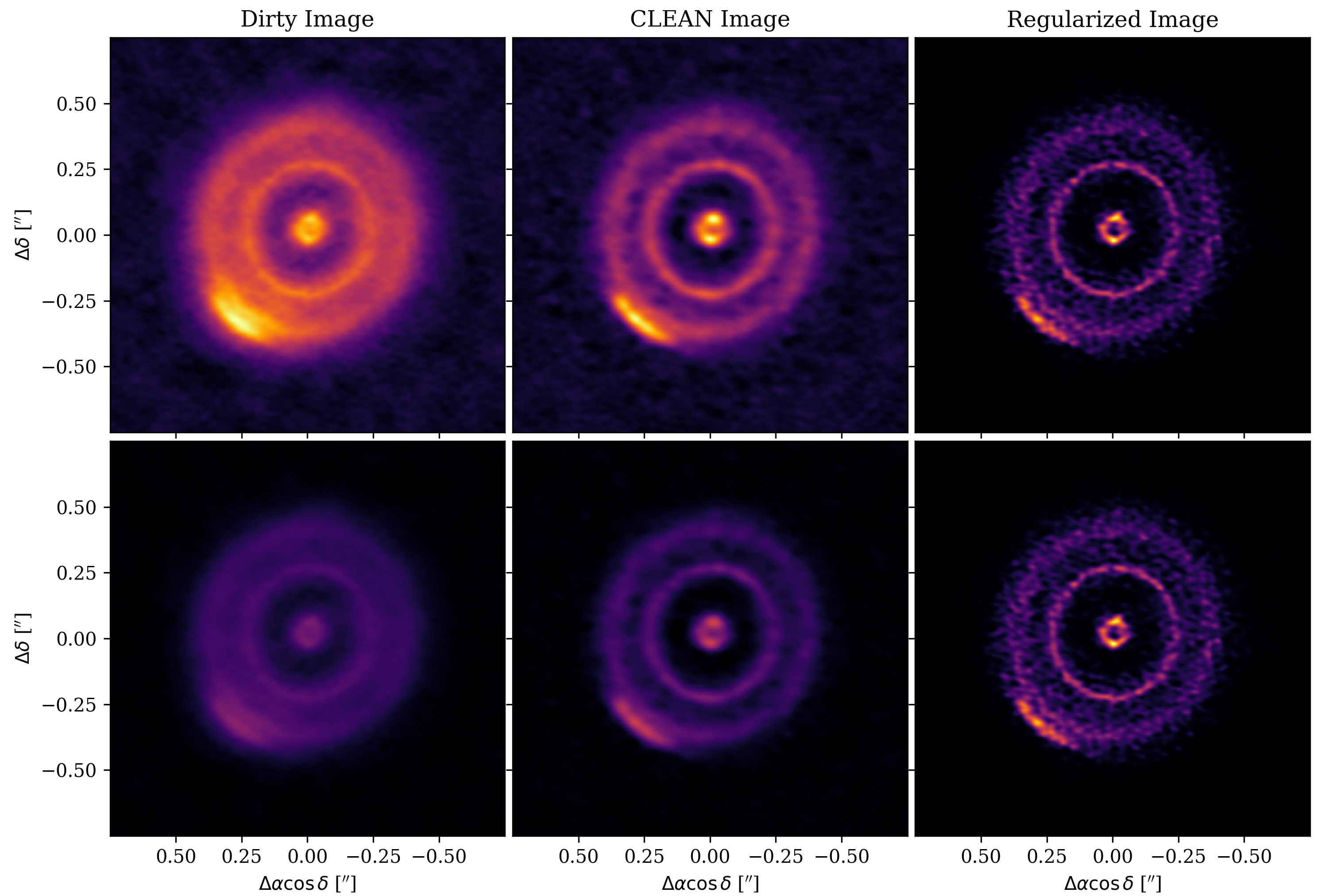}
\caption{Dirty image (left, robust=0.5), \texttt{tclean} from \citet{Andrews_2018} (center), and RML (right) images of HD~143006. The top row shows the three images on individual color scales (i.e. each panel is normalized to the minimum and maximum pixel values), while the bottom row shows the three images displayed on the same color scale. The loss function for the RML image included maximum entropy ($\lambda=1 \times 10^{-1}$), sparsity ($\lambda=5 \times 10^{-4}$), and TSV ($\lambda=2.15 \times 10^{-3}$) terms. In this case, maximum entropy contributes high resolution features to the image, sparsity removes background noise, and TSV helps better define ring structure in the disk.}
\label{fig:clean_compare}
\end{figure*}

We distilled the practical wisdom gained from our exploration of regularizers and image validation to produce an exemplary image from the HD~143006 dataset. We combined entropy, sparsity, and TSV regularization to generate an image of HD~143006, shown in Figure \ref{fig:clean_compare}. Entropy contributed high resolution features to the image, while sparsity removed background noise. We found that TSV helps better define ring structure in the disk, but excluded TV because it degraded the resolution of the image. We also provide a comparison to the CLEAN image synthesized by the DSHARP team \citep{Andrews_2018}. To tune the hyperparameters, we used random cell CV to identify a starting point, then adjusted $\lambda_{\rm{TSV}}$ by hand to introduce more smoothing into the model image. We increased the $\lambda_{\rm{TSV}}$ from $2.15 \times 10^{-4}$ to $2.15 \times 10^{-3}$, a change comparable to the range of acceptable hyperparameter values identified in Figures \ref{fig:cv_variation} and \ref{fig:dartboard_vs_randcell}. We found that RML methods successfully recover all features that appear in the fiducial CLEAN image of HD~143006 synthesized by the DSHARP team \citep{Andrews_2018}, as well as remove background noise and improve angular resolution compared to the CLEAN image.

\section{Discussion} \label{sec:discussion}

\subsection{Best Practices for Image Validation}

Cross-validation (CV) is a tool for determining the settings of the regularization parameters that yield an image model with the best predictive power for new data. The use of CV methods on interferometric data is an active area of research, and there are many aspects which have not yet been fully explored. First, CV can be implemented as either an exhaustive or non-exhaustive method; an exhaustive method will use all possible ways to partition data, while a non-exhaustive method will use only a subset of possible partitions. An example of an exhaustive method is leave-one-out CV (LOOCV), which takes all but one data point as the training set and uses the remaining data to test, cycling through the entire dataset. These methods can be extremely computationally expensive, especially in the case of interferometric data containing millions of visibility measurements. In addition, LOOCV performs worse in terms of parameter selection and evaluation compared to other methods of CV \citep{Breiman_1992}.

Non-exhaustive CV methods like K-fold CV greatly reduce the total computational burden, and are thus a more practical CV method for interferometric data. K-fold CV requires a choice of $K$ that balances bias and variance in the parameter error estimates, with a high $K$ yielding a low bias, high variance estimation and a low $K$ yielding a high bias, low variance estimation \citep{Hastie_2009}. Studies in statistics and informatics have consistently found $K=10$ to provide the best bias-variance trade-off \citep[e.g.][]{Breiman_1992, Kohavi_1995, Molinaro_2005}. While no studies have specifically examined the optimal $K$ for K-fold CV of interferometric data, the standard $K=10$ has been used with success for such applications \citep{Akiyama_2017a, Akiyama_2017b, Yamaguchi_2020}. In this study, we restricted our exploration of CV to K-fold CV with $K=10$.

Aside from the choice of $K$, one must also decide how to partition the data set. Random cell partitioning randomly selects visibility grid cells without replacement for each subset of testing data. With such a large number of visibility grid cells, it is likely that each subset will have similar $(u,v)$ coverage. Therefore, CV with random cell partitioning effectively tests how well a trained model predicts new data with similar $(u,v)$ coverage. Random cell partitioning has been used to tune regularizer hyperparameters in previous studies of RML for interferometry \citep[e.g.][]{Akiyama_2017a, Akiyama_2017b, Yamaguchi_2020}. We also explore dartboard partitioning, which generates testing data from radial and azimuthal bins of gridded visibility cells (see Figure \ref{fig:kfold_mask}). CV with dartboard partitioning tests how well the model extrapolates to $(u,v)$ space notably different than the training data. Dartboard partitioning aims to simulate the irregular $(u,v)$ sampling common to most ALMA interferometric observations (e.g., across execution blocks or array configurations), approximating how the model might fit with data obtained from a variety of array configurations. As the number of dartboard bins increases, the dartboard partitioning scheme begins to test the predictive power of the model in comparable u-v space (as with random cell partitioning). CV partitioning schemes for interferometric imaging are an active area of research, but we find that both random cell and dartboard methods are helpful in determining the range of performant hyperparameter values. 

\begin{figure*}
\centering
\includegraphics[width=0.94\linewidth]{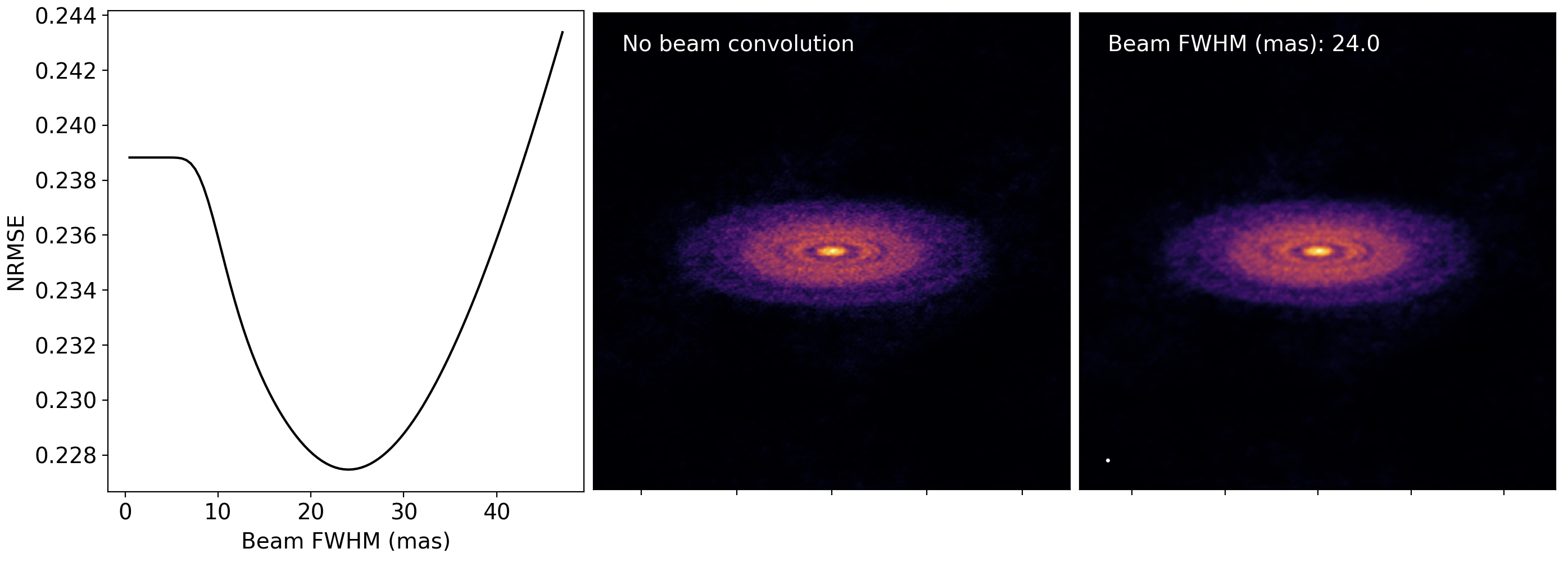}
\caption{Convolving an RML image with a beam that minimizes the NRMSE between the RML image and the original simulation. The left panel shows how the NRMSE changes with beam FWHM values. The NRMSE is minimized at a beam FWHM of 24.0 mas. The center panel shows the RML image of the simulated disk dataset ($\lambda_{\rm{entropy}} = 2.15 \times 10^{-1}$, $\lambda_{\rm{TSV}} = 4.64 \times 10^{-3}$, $\lambda_{\rm{sparse}} = 2.00 \times 10^{-3}$). The right panel shows this RML image convolved with a 24.0 mas restoring beam.
}
\label{fig:nrmse_minimized}
\end{figure*}

Regardless of the choice of $K$ and data partitioning scheme, it is vital to ensure that the model has fully converged for each training set. We find that the number of iterations needed to optimize the model during K-fold CV is often greater than the number of iterations needed to optimize the model with the full dataset, as training on fewer data points can require more iterations before the loss function is minimized. This is especially true for the K-fold that does not contain the visibilities at the lowest spatial frequencies (i.e. $(u,v)$ close to zero), as the omission of this data can cause a slow initial decline of the total loss. We found that for dartboard partitioning, the training set without the lowest spatial frequency visibilities is usually the slowest to converge.

Because the final CV score is the sum of the $\chi^2$ for all K-folds, a K-fold that has not fully converged can result in a spuriously high CV score. We recommend inspecting each K-fold for convergence after training, as well as inspecting the final $\chi^2$ value for each K-fold. This delayed convergence usually coincides with the K-fold for which the training set lacks data at the lowest spatial frequencies (e.g. K-fold 5 in Figure \ref{fig:kfold_mask}), and commonly occurs when using dartboard visibility partitioning. If each K-fold does not reach convergence, the final CV score is invalid and the entire CV process must be repeated with enough iterations to ensure full convergence.

While CV scores can be used to find optimal regularization parameters, it is important to take care when comparing CV scores. First, as mentioned above, any CV score obtained from training without allowing the loss function to reach a minimum (i.e. one or more K-folds do not reach convergence) cannot be used. CV scores may only be compared across the same dataset, model specification, and CV setup. The CV setup includes the value of $K$ and the choice and implementation of partitioning scheme.

It is possible to compare CV scores when using different regularizers, so long as the regularizer has a tuneable prefactor (e.g. $\lambda$ values) that can be set to zero. For example, the CV score for a model with only an entropy regularizer can be straightfowardly compared to that of a model with only a TV regularizer, because this is effectively comparing different ways to tune the $\lambda$ prefactors, including $\lambda_{\rm{entropy}}=0$ and $\lambda_{\rm{TV}}=0$. In addition to these edge cases which ``turn off'' certain regularizers, CV scores can be compared for any values of $\lambda_{\rm{entropy}}$ and $\lambda_{\rm{TV}}$, as long as the dataset and CV setup remain consistent.

\subsection{Determining Image Resolution with RML}

\begin{figure*}
\centering
\includegraphics[width=0.94\linewidth]{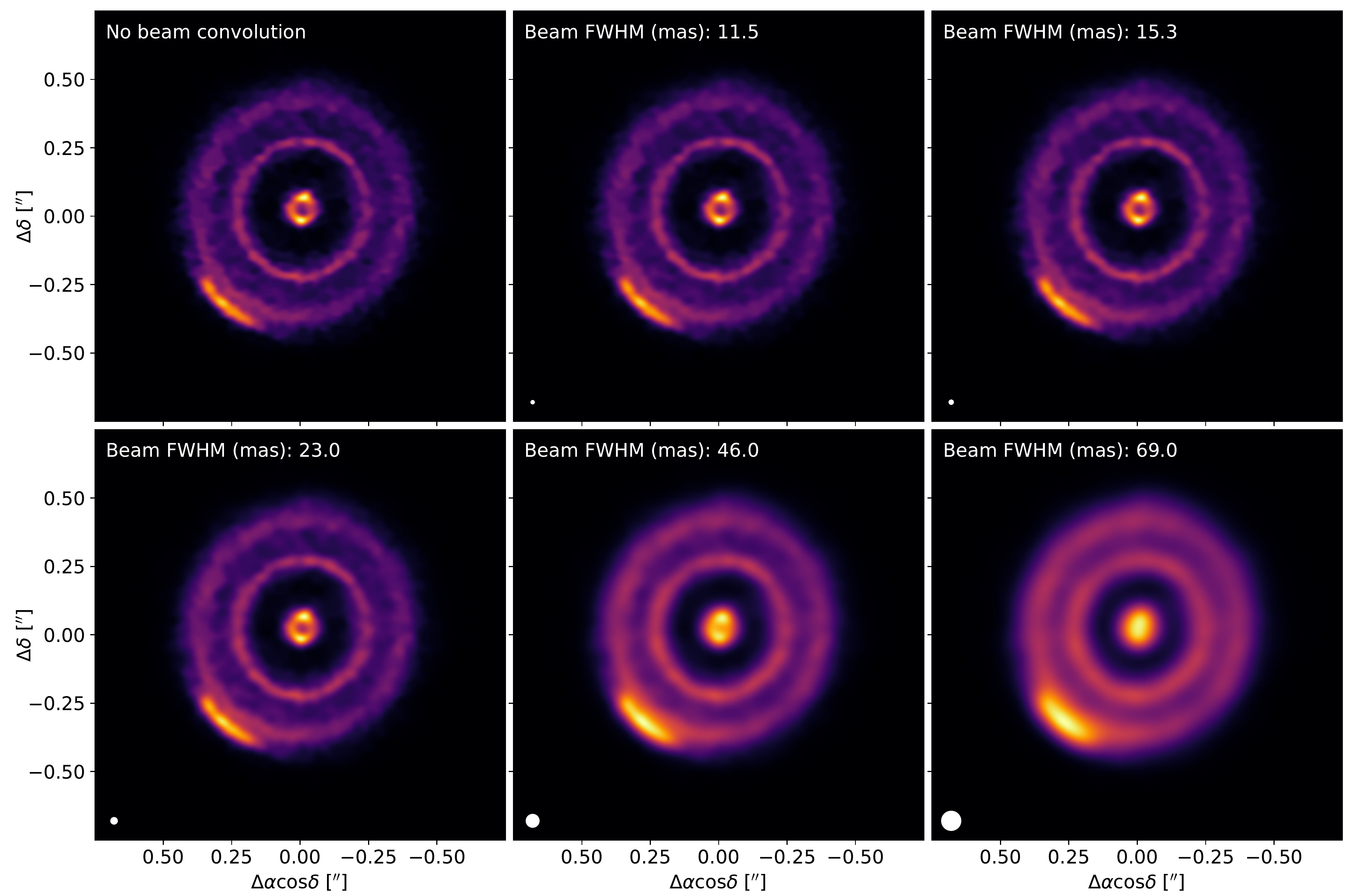}
\caption{The effect of a restoring beam on an RML image of HD~143006. As in Figure \ref{fig:clean_compare}, the loss function for the RML image included maximum entropy ($\lambda=1 \times 10^{-1}$), sparsity ($\lambda=5 \times 10^{-3}$), and TSV ($\lambda=2.15 \times 10^{-3}$) terms. The top left panel shows the RML image with no restoring beam. All other panels show the same image after being convolved with a circular Gaussian restoring beam, with beam sizes shown in the bottom left corner of each panel. The beam sizes shown are based off the resolution of the CLEAN image made from the DSHARP data of HD~143006, which has a synthesized beam FWHM of $45 \times 46$ mas \citep{Andrews_2018, Perez_2018}. We use this as a reference resolution, and show the RML image restored to 46 mas in the bottom center panel. We also show images convolved with beams of 1/4 (top center), 1/3 (top right), 1/2 (bottom left), and 3/2 (bottom right) the size of the 46 mas reference.
}
\label{fig:beam_convolution}
\end{figure*}

Imaging with CLEAN typically involves building up a model of CLEAN components and then convolving that model with the CLEAN beam. A CLEAN component may be as simple as a Dirac $\delta$-function, which is useful for fields with many point sources, but may be the suboptimal basis set for representing spatially resolved sources. Beam convolution effectively spreads flux from these components over the size of the beam, making the image more representative of the true source morphology at the cost of resolution. Because the size of the CLEAN beam is a known quantity, characterizing the resolution of a CLEANed image is relatively straightforward.

RML images, on the other hand, are not generated from a set of individual components and thus do not require beam convolution in order to obtain a smoother image product. The most obvious benefit to this is that a strict limit on resolution is not baked into the imaging workflow. However, the lack of restoring beam does make characterizing the resolution of an RML image more ambiguous than characterizing the resolution of a CLEAN image. \citet{Chael_2016} find that restoring beams can still be useful for RML methods, as false high-frequency features can sometimes be present in the image. However, we find that carefully selecting and tuning regularizers is a more effective way to ensure that erroneous features do not appear, as a restoring beam that is too large could remove real features in the image.

The theoretical restoring beam size can be computed given knowledge of the true source; the beam size that minimizes the NRMSE is commonly adopted as the primary metric for evaluating the quality of reconstructed interferometric images \citep[e.g.][]{Chael_2016,Akiyama_2017a,Akiyama_2017b, Kuramochi_2018, Yamaguchi_2020}. Notably, \citet{Chael_2016} show that when trying to recover a ``true'' reference model input image of a compact source, the NRMSE is minimized at a considerably smaller beam size with RML techniques compared to CLEAN. If the ground truth is known (e.g. when making an RML image from simulated data based off of a model image, as in our simulated disk dataset), then convolving the RML image with a beam size that minimizes the NRMSE can maintain the highest degree of superresolution in the image while removing any potential spurious high-frequency features.

Figure \ref{fig:nrmse_minimized} shows an RML image convolved with a circular Gaussian restoring beam that minimizes the NRMSE. We use the simulated disk dataset (so that we have a ground truth for comparison) and the RML model tuned by random cell CV. The base RML model and the beam convolved RML model look similar, with the only difference being a small degree of blurring/smoothing in the convolved image. Because the NRMSE at a relatively small beam size (FWHM = 24.0 mas), it is unsurprising that the images do not deviate from each other significantly. Given that perfect knowledge of the true sky brightness does not accompany observational data, this approach will not always be possible. Even though the optimal restoring beam size cannot always be constrained, image fidelity is only mildly worsened by selecting a beam size smaller than the optimal value. As a result, the use of a restoring beam in an RML imaging workflow can only result in small fidelity gains in the best case, and can result in significant loss of resolution in the worst case.

RML methods are known to be capable of generating images superresolved to 1/4 of the nominal resolution of the interferometer $R_{\min }=\lambda / b_{\max}$, where $b_{\max}$ is the length of the longest baseline in the array \citep[e.g.][]{Narayan_1986,Honma_2014}. This is typically treated as an upper resolution limit for RML methods, as the derivation of this factor assumes that the data have high signal-to-noise and thoroughly sample the visibility function \citep{Holdaway_1990}. In practice, superresolution factors ranging from roughly 1/3 to 1/2 of the nominal resolution are the most common outcome from RML imaging methods \citep[e.g.][]{Chael_2016, Akiyama_2017b, Akiyama_2017a, Cieza_2017, Casassus_2018, Kuramochi_2018, Casassus_2019, Casassus_2021}.

Figure \ref{fig:beam_convolution} shows RML images of HD~143006 both without any restoring beam, and with restoring beams equal to 1/4, 1/3, 1/2, 1, and 3/2 times the synthesized beam size of the DSHARP continuum \texttt{tclean} image of HD~143006 (11.5, 15.3, 23.0, 46.0, and 69.0 mas, respectively). We used \texttt{astropy.convolution} to convolve the model with circular Gaussians directly in the image plane. There are no significant qualitative differences in the base RML image (top left panel) and the RML images convolved with beams equal to 1/4 (top center), 1/3 (top right), and 1/2 (bottom left) the size of the synthesized beam of the \texttt{tclean} image. In this case, we do not observe spurious high-frequency features in the base RML image, so convolution with small restoring beams has little impact on the final image. We emphasize that while RML methods do not require a restoring beam, convolving an RML image with a modest restoring beam (i.e. 1/3-1/2 the nominal resolution of the observations, consistent with the performance seen in Figures \ref{fig:nrmse_minimized} and \ref{fig:beam_convolution}) yields a more conservative final image while still benefitting from some degree of superresolution.


\section{Conclusion} \label{sec:conclusion}

We have developed \texttt{MPoL}, a GPU-accelerated RML imaging package for image synthesis of complex visibilities from ALMA. We described the mathematical foundation of RML imaging methods and several regularizers, and described a general RML imaging framework with \texttt{MPoL}. We explored how maximum entropy, sparsity, TV, and TSV regularizers can be incorporated into the imaging process, and how each of these regularizers impacts image synthesis of ALMA continuum data of protoplanetary disk datasets. We found that for both real data of the HD~143006 protoplanetary disk and simulated protoplanetary disk data, a combination of entropy, sparsity, and TSV regularization works well, while TV regularization does not adequately retain fine details in the images. With these methods we improved the angular resolution of the images by a factor of $2-3$ compared to CLEAN.

In addition, we explored CV methods as a robust procedure for hyperparameter tuning and image validation to maximize image fidelity and resolution. We tested K-fold CV with random cell visibility partitioning and novel dartboard partitioning, comparing these methods to tuning hyperparameters by trial and error. We found that tuning by random cell CV or by eye achieved images closest to the ground truth, while dartboard partitioning resulted in a similar but slightly noisier image. We found that a range of hyperparameter values can result in comparably low CV scores, suggesting that it is not necessary to precisely tune hyperparameters according to the CV score (which can impose a computational burden). Rather, using CV across a coarse grid of hyperparameter values is an efficient way to guide the tuning process.

Overall, RML techniques provide flexible imaging processes that are well-suited for applications to ALMA continuum protoplanetary disk measurement sets. The use of RML techniques can improve image fidelity on small scales, and expanding applications to spectral line data has the potential to aid the detection and characterization of kinematic disturbances within protoplanetary disks (such as the discovery of the circumplanetary disk candidate in the disk around AS~209 presented in \citet{Bae_2022}). Exploring regularization on a wider range of source morphologies observed by ALMA, including image cubes with many channels, will broaden our understanding of how image synthesis with RML techniques can benefit different science cases.

\begin{acknowledgments}
We acknowledge funding from an ALMA Development Cycle 8 grant number AST-1519126. This paper makes use of the following ALMA data: ADS/ JAO.ALMA\#2016.1.00484.L. ALMA is a partnership of ESO (representing its member states), NSF (USA) and NINS (Japan), together with NRC (Canada), MOST and ASIAA (Taiwan), and KASI (Republic of Korea), in cooperation with the Republic of Chile. The Joint ALMA Observatory is operated by ESO, AUI/NRAO and NAOJ. The National Radio Astronomy Observatory is a facility of the National Science Foundation operated under cooperative agreement by Associated Universities, Inc. Yina Jian is a summer student at the National Radio Astronomy Observatory. Computations for this research were performed on the Pennsylvania State University’s Institute for Computational and Data Sciences’ Roar supercomputer. The Center for Exoplanets and Habitable Worlds is supported by the Pennsylvania State University, the Eberly College of Science, and the Pennsylvania Space Grant Consortium. This research has made use of NASA’s Astrophysics Data System Bibliographic Services. We thank Christophe Pinte for providing the protoplanetary disk simulation used to make our synthetic data. We thank Poe for contributing his likeness to Figure 2.
\end{acknowledgments}

\software{Astropy \citep{astropy_2013, astropy_2018}, PyTorch \citep{PyTorch_2019}, MPoL \citep{mpol_2021}, CASA \citep{McMullin_2007}}

\appendix 
\section{Recommendations for RML with ALMA} \label{app:recs}
The \texttt{MPoL} software used in this study is open source and designed with ALMA measurement sets in mind. We recommend that anyone wishing to use \texttt{MPoL} for RML imaging of ALMA data follow these general steps.
\begin{enumerate}
    \item {\bf Obtain arrays of complex visibility data.} \texttt{MPoL} is designed to work directly with arrays of complex visibilities. Visibilities can be obtained from CASA measurement sets using \texttt{casatools}. Open source software packages like visread\footnote{https://mpol-dev.github.io/visread/} can aid in this process.
    \item {\bf Select pixel size and number of pixels in the image.} These pixels will serve as the model parameterization, and allow ungridded visibilities to be gridded. It is important to be mindful of the choice of pixel size, as pixels that are too large will place an intrinsic resolution limit on the final image, while using too many small pixels will introduce an unnecessary computational burden during the imaging process. At this point, it is also a good idea to create a dirty image with \texttt{MPoL} to make sure that everything has been loaded and initialized as expected.
    \item {\bf Set the initial state of the model and determine which regularizers to include in the loss function.} We recommend initializing the model with the dirty image. Dirty image initialization leads to faster convergence of the loss function. The loss function may include any number of regularizing terms.
    \item {\bf Define a range of hyperparameter values to be tested with CV.} Tuning hyperparameters with CV is the most time-consuming part of this imaging workflow, as each hyperparameter setting is tested on multiple subsets of the full dataset (the training data). The total amount of time can be minimized by carefully selecting the range of hyperparameter values to be tested. We recommend beginning with a coarse grid of values. The hyperparameter values that yield the highest fidelity images can change with the dataset, so starting with a wide range of values (i.e. spanning several orders of magnitude) helps quickly hone in on a narrower range of potential values. For example, while one dataset might minimize the CV scores with an entropy prior with $\lambda=0.1$, another dataset might minimize the CV score with $\lambda=10$, so a coarse round of CV might include $\lambda=[0.01,0.1,1,10]$.
    
    Because generating an image is computationally much faster than running the full CV process, visualizing the effects of different regularizer strengths on the dataset (e.g. as in Figure \ref{fig:image_grid}) is an efficient way to determine the initial range of hyperparameter values to be tested with CV. While synthesizing images with the full dataset is not necessary at this stage, the extra effort is useful for constraining the range of plausible hyperparameter values and lowers the possibility of testing hyperparameter values that result in a poor fit to the data. Additionally, datasets which share features like source morphology and baseline coverage (like the two disk datasets presented in this study) may have optimal hyperparameter values that are similar, so hyperparameter values used for imaging comparable datasets could provide a good starting point for the CV process, though this will not always be possible.
    \item {\bf Set up the CV process by defining the number of K-folds and data partitioning scheme.} We recommend using $K=10$, which has been shown to balance bias and variance in the estimation and has already been used with success in interferometric imaging.
    \item {\bf Perform CV (perhaps in a coarsely-defined round and a fine-tuning round) to obtain optimal imaging hyperparameters.} The set of hyperparameters that minimizes the CV score corresponds to the model with the best predictive performance. Ensure that each K-fold has reached convergence during the CV process; if each K-fold did not fully converge, the CV score is invalid and CV must be restarted with enough iterations to allow full convergence. For the HD~143006 dataset, computing a single CV score ($K=10$, 15000 iterations each) took around 11-12 minutes on a NVIDIA Tesla K80 GPU. This will scale depending on the number of hyperparameter combinations tested and the number of GPUs employed. For example, performing CV for a $7\times7$ grid of $\lambda$ combinations (such as those shown in Figure \ref{fig:dartboard_vs_randcell}) took roughly 2.5 hours on 4 GPUs, though on a single GPU this would increase to nearly 9.5 hours. If CV is performed in multiple rounds, then smaller (and thus faster) grids are practical.
    \item {\bf Image using the full dataset with the hyperparameters that minimized the CV score.} Ensure that the model has fully converged. If it has, the result is a fully regularized image. At this stage, it may be useful to adjust hyperparameters by hand and assess the changes visually to determine whether further tuning is needed. For the HD~143006 dataset, generating a single image took about 10 seconds NVIDIA Tesla K80 GPU.

    {\bf }
    \item {\bf Optionally, convolve the image with a restoring beam.} While beam convolution is not required in an RML imaging workflow, some users may find it helpful for characterizing the resolution of the image by filtering out potential spurious high-resolution features in the image. We recommend restoring the image to no more than 1/3 to 1/2 of the nominal resolution in order to retain the superresolution benefits of RML imaging.
    
\end{enumerate}

While this is a rough outline of a functional RML workflow, in practice the RML imaging process need not be so linear. For some imaging cases, one may find it beneficial to try individual regularizers before combining them, or to produce some preliminary images by hand tuning regularizers before running the full CV process. When selecting regularizers, we recommend considering the qualitative features the source is likely to have. Maximum entropy and sparsity regularization tend to produce high-resolution features, and sparsity can effectively remove background noise. TV and TSV regularizers tend to emphasize sharp edges in the image, with TV enforcing sharp edges more rigidly. These general characteristics can help inform which regularizers to include in the loss function.

\bibliography{references}{}
\bibliographystyle{aasjournal}

\end{document}